\RequirePackage{fix-cm}
\PassOptionsToPackage{sort}{natbib}
\documentclass[smallextended,natbib]{svjour3}
\smartqed
\usepackage{amsmath,amssymb,amsfonts}
\usepackage{graphicx}
\usepackage{textcomp}
\usepackage{multirow}
\usepackage{pifont}
\usepackage{booktabs}
\usepackage{xcolor}
\usepackage{adjustbox}
\usepackage{tcolorbox}
\usepackage[paperwidth=155mm, paperheight=235mm, textwidth=11.9cm, textheight=19.4cm, includehead, centering]{geometry}
\usepackage[colorlinks=true, urlcolor=blue, linkcolor=black, citecolor=black]{hyperref}
\usepackage{orcidlink}
\newtcolorbox{answerbox}{colback=gray!8, colframe=gray!55!black, boxrule=0.5pt, arc=1pt, left=4pt, right=4pt, top=3pt, bottom=3pt}
\newenvironment{declitem}[1]{\par\addvspace{6pt}\small\rmfamily\trivlist\item[\hskip\labelsep{\bfseries #1}]}{\endtrivlist}
\newcommand{\tabnotes}[1]{\par\vspace{3pt}\begin{minipage}{\textwidth}\footnotesize #1\end{minipage}}

\newcommand*{\doilink}[1]{\url{https://doi.org/#1}\endgroup}

\renewcommand{\makeheadbox}{}

\begin{document}

\title{PatchCon: Patch-aware Weighted Supervised Contrastive Learning for Vulnerability Detection}
\titlerunning{PatchCon: Patch-aware Weighted Supervised Contrastive Learning}

\author{Chaomeng Lu~\orcidlink{0000-0003-1841-2698} \and
        Pooya Rostami Mazrae~\orcidlink{0000-0002-4859-1546} \and
        Bert Lagaisse~\orcidlink{0000-0002-2543-6658}
}
\authorrunning{Chaomeng Lu et al.}

\institute{Chaomeng Lu (corresponding author) \and Pooya Rostami Mazrae \and Bert Lagaisse \at
              DistriNet Group T, KU Leuven, Leuven, Belgium \\
              \email{chaomeng.lu@kuleuven.be; pooya.rostamimazrae@kuleuven.be; bert.lagaisse@kuleuven.be}
}

\date{}

\maketitle

\begin{abstract}
Automated vulnerability detection is critical for securing modern software systems, yet existing learning-based approaches often treat functions as independent samples and underuse the supervision provided by vulnerability-fixing patches. In this paper, we propose PatchCon, a CodeBERT-based patch-aware contrastive learning model for function-level vulnerability detection. PatchCon jointly optimizes vulnerability classification and supervised contrastive learning, and treats vulnerable--patched function pairs as hard negatives with larger contrastive weights to capture subtle security-relevant differences introduced by fixes. To evaluate patch-level discrimination, we use two pairwise metrics that measure whether a model can correctly distinguish vulnerable functions from their patched counterparts. We evaluate PatchCon on PairVul-Linux through extensive experiments, including state-of-the-art comparisons, ablation studies, hyperparameter sensitivity analysis, and statistical significance testing. We also conduct fine-grained analyses by Common Weakness Enumeration (CWE) type, function length, and patch size, and further evaluate PatchCon on Linux kernel vulnerabilities disclosed after the training data cutoff. On PairVul-Linux, PatchCon outperforms representative pretrained code models and contrastive learning baselines on both classification and pairwise metrics, improves pairwise accuracy over the strongest reproduced baseline from 0.2060 to 0.3264, and achieves pairwise accuracy comparable to that of a GPT-4o-based large language model (LLM) method at a substantially lower cost. Among the reproduced models, PatchCon also achieves the highest pairwise accuracy on test pairs from Common Vulnerabilities and Exposures (CVE) entries unseen during training and on newly disclosed vulnerabilities.
\end{abstract}

\section{Introduction}
Software vulnerabilities remain a persistent threat to modern software ecosystems~\citep{ohm2020backstabber}. As large-scale open-source repositories are increasingly reused in safety-critical, financial, governmental, and cloud infrastructures, vulnerable code can propagate rapidly across projects and affect downstream users at scale~\citep{ladisa2023sok}. Automated vulnerability detection has therefore become an important research direction in software security~\citep{yamaguchi14modeling,le2022survey}. Recent approaches based on deep learning and Large Language Models (LLMs) have shown promise in identifying vulnerable functions from source code, but reliably distinguishing vulnerable code from its patched counterpart remains challenging because vulnerability fixes often involve subtle yet security-critical changes~\citep{steenhoek23an,zhou25large,niklas24uncover}.

Existing vulnerability detection approaches still face three major limitations.

\emph{Unreliable datasets.} The datasets used in many previous studies are often not sufficiently reliable for realistic and fine-grained vulnerability detection evaluation. 
Some commonly used datasets suffer from low label accuracy, high duplication rates, and a large number of functions that are not directly related to vulnerabilities. 
In other cases, the data may not be constructed from real-world vulnerability reports and corresponding patches, making it difficult to assess whether a model can generalize to practical vulnerability detection scenarios~\citep{croft2022data,chen23diverse,primevul25}.

\emph{Underused patch supervision.} Many existing methods formulate vulnerability detection as an independent single-function classification task. However, real-world vulnerabilities are often best understood through their fixes. A vulnerable function and its patched counterpart usually share most lexical and syntactic structures, while differing only in small but security-critical edits. Such vulnerable--patched pairs provide strong supervision signals for learning security-sensitive representations, but this information has not been fully exploited in many existing pipelines~\citep{tan25similar}. Although contrastive learning has recently been introduced into this field~\citep{cao2025mcl,tian26when}, existing designs either do not use vulnerable--patched pairs or treat them in the same way as ordinary negative samples.

\emph{Limited patch-level evaluation.} Existing evaluation protocols provide limited insight into patch-level discrimination. Most detectors are still evaluated only with standard classification metrics, such as accuracy, precision, and F1-score. These metrics are useful for measuring overall detection performance, but they do not directly evaluate whether a model assigns higher vulnerability confidence to a vulnerable function than to its fixed version. This ability is critical in practice, because a reliable vulnerability detector should be able to distinguish a vulnerable implementation from a minimally changed but security-corrected patch.

To address these limitations, this paper proposes \textbf{PatchCon}, a \textbf{Patch}-aware Weighted Supervised \textbf{Con}trastive Learning framework for function-level vulnerability detection. To support reliable and fine-grained evaluation, we conduct experiments on PairVul-Linux~\citep{du26vulrag}, which is constructed from real-world Linux kernel Common Vulnerabilities and Exposures (CVE) reports and their corresponding patches, removes functions that are not directly related to vulnerabilities, and applies strict filtering to improve label quality. In addition, we use the Linux kernel subset of VentiVul~\citep{lu2025from}, a test set of vulnerabilities published more than a year after the most recent vulnerability in PairVul-Linux, to evaluate whether the learned discrimination generalizes to newly disclosed vulnerabilities. 

Different from conventional single-function classifiers, PatchCon explicitly incorporates patch-pair information by constructing four types of function pairs. 
In particular, vulnerable--patched pairs are treated as hard negatives and assigned larger contrastive weights. 
This design encourages the model to learn compact intra-class representations while enlarging the representation gap between vulnerable functions and their patched counterparts. 
Beyond standard classification metrics, we use two pairwise evaluation metrics, Pairwise Accuracy (PACC), which follows the pairwise evaluation of prior work~\citep{du26vulrag}, and Mean Confidence Gap (MCG), to directly evaluate whether a model ranks vulnerable functions as more vulnerable than their corresponding fixed versions.
We further conduct a comprehensive evaluation around six research questions (RQs) and discuss the robustness of the results to CVE overlap and representative cases.
The main contributions of this paper are summarized as follows:

\begin{itemize}
    \item We propose PatchCon, a supervised contrastive approach to function-level vulnerability detection that treats vulnerable--patched function pairs as hard negatives, weights them more heavily than ordinary negatives, and is trained jointly with classification.
    \item We evaluate PatchCon on PairVul-Linux and VentiVul with standard and pairwise metrics. PatchCon improves PACC over the strongest reproduced baseline from 0.2060 to 0.3264, with statistically significant gains over all evaluated baselines, and achieves pairwise accuracy comparable to a GPT-4o-based method at a much lower cost.
    \item We show that the advantage of PatchCon in vulnerable--patched discrimination persists without CVE overlap: it retains the highest PACC among the reproduced models on test pairs from CVEs unseen during training and on VentiVul.
\end{itemize}

This paper is structured as follows. Section~\ref{sec:related_work} reviews related work on vulnerability detection, contrastive learning, and patch-aware vulnerability analysis. Section~\ref{sec:methodology} presents the design of PatchCon. Section~\ref{sec:experimental_setup} describes the datasets, baselines, implementation details, and evaluation metrics. Section~\ref{sec:evaluation} reports the results for the six RQs. Section~\ref{sec:discussion} discusses the robustness of the results to CVE overlap, a case study, limitations, and threats to validity. Section~\ref{sec:conclusion} concludes the paper.

\section{Related Work}
\label{sec:related_work}
Early software vulnerability detection approaches relied on handcrafted features, static analysis rules, or symbolic reasoning~\citep{yamaguchi14modeling,ghaffarian18software}. With the advent of neural code models, vulnerability detection has shifted toward data-driven representation learning~\citep{vuldeepecker18,devign19,sysevr22}. More recently, programming language models and LLMs have been fine-tuned or prompted for vulnerability classification and have shown promising results on multiple benchmarks~\citep{wen24livable,shiri24a,sheng26large}. However, recent studies also show that learning-based detectors can be sensitive to dataset artifacts and distribution shifts, and may struggle to distinguish vulnerable functions from semantically similar patched counterparts~\citep{chakraborty24revi,croft23data,lu2025from}. These findings motivate vulnerability detection methods that explicitly exploit patch-level supervision rather than treating functions only as independent samples.

Contrastive learning (CL) has become a powerful representation learning paradigm through both self-supervised and supervised objectives~\citep{He2020CVPR,chen20apmlr,supcon20}. In software engineering, contrastive objectives have been explored for clone detection, defect prediction, code search, and code representation pretraining~\citep{contrabert23,shi23cocosoda,xu23contrast}. For vulnerability detection, 
several studies have adapted CL to improve representation robustness and vulnerability-specific discrimination.
ReVeal~\citep{chakraborty22deep} represents source code with graph-based features and introduces triplet-loss-based CL for vulnerability detection. ContraFlow~\citep{chen22path} leverages self-supervised CL to preserve feasible value-flow paths, improving vulnerability classification and buggy-statement localization. ContraBERT~\citep{contrabert23} combines masked language modeling with contrastive pretraining over augmented code and natural-language variants to improve pretrained code representations.
Clear~\citep{chen24clear} applies CL to smart contract vulnerability detection by modeling fine-grained correlations among contracts. \citet{ji24applying} propose a hierarchical CL framework for long-tail multi-class vulnerability classification by leveraging relationships among Common Weakness Enumeration (CWE) classes and max pooling for long-code representation. SCL-CVD~\citep{wang24sclcvd} applies supervised contrastive learning to vulnerability detection by fine-tuning GraphCodeBERT with Supervised Contrastive Learning (SupCon) loss. MCL-VD~\citep{cao2025mcl} combines source code, comments, and abstract syntax tree (AST) information with multimodal CL to improve vulnerability detection. 
These approaches demonstrate the usefulness of CL for vulnerability-related representation learning, but most of them optimize general code representations or function-level class separation without explicitly modeling vulnerable--patched functions as hard negative pairs.

Several works are more closely related to patch-aware or repair-aware vulnerability analysis. VDSimilar~\citep{sun21vdsimilar} exploits vulnerable--patched pairs with a Siamese BiLSTM-Attention network to improve vulnerability detection under limited labeled data. DELTA~\citep{tian26when} uses vulnerable--patched function pairs and optimization-varied binaries as contrastive samples for repair-aware binary vulnerability detection. Vul-RAG~\citep{du26vulrag} enhances LLM-based vulnerability detection with retrieval-augmented knowledge distilled from historical vulnerabilities and fixes, improving the ability of LLMs to distinguish vulnerable code from similar patched code. These studies highlight the importance of patch information, but they differ from PatchCon in either task setting, model design, or evaluation focus. In particular, DELTA targets binary code and uses vulnerable--patched pairs in an NT-Xent objective without pair-type weighting. In contrast, PatchCon targets source-code function-level vulnerability detection, assigns larger weights to vulnerable--patched pairs than to ordinary negatives in a supervised contrastive objective jointly optimized with classification, and uses pairwise metrics to directly evaluate patch-level vulnerable--patched discrimination.

\section{Methodology}
\label{sec:methodology}

\begin{figure}[ht]
    \centering
    \includegraphics[width=1\linewidth]{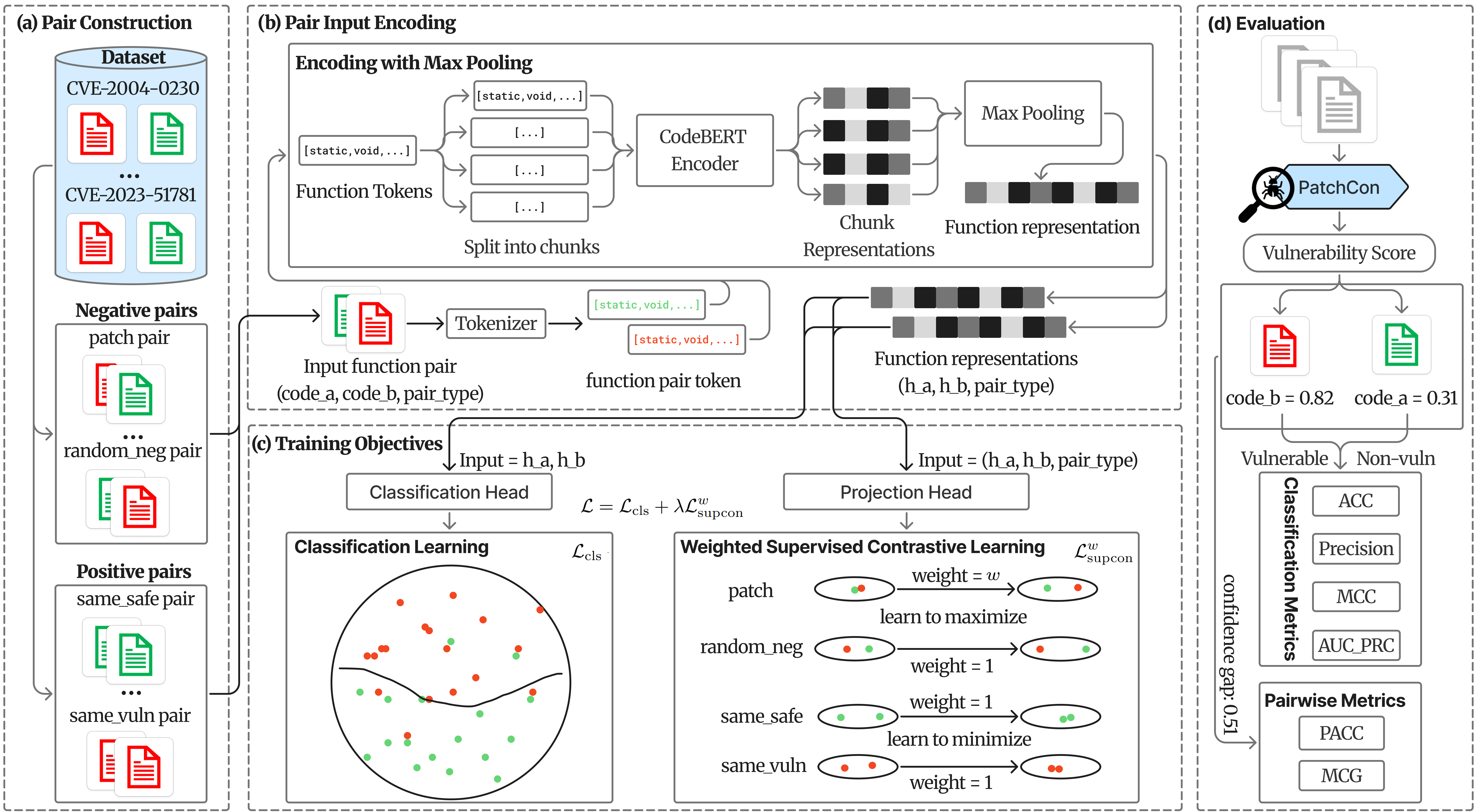}
    \caption{Overview of the proposed PatchCon framework.}
    \label{fig:architecture}
\end{figure}

As shown in Fig.~\ref{fig:architecture}, PatchCon consists of four components: (a) Pair Construction, (b) Pair Input Encoding, (c) Training Objectives, and (d) Evaluation. Given vulnerability-fixing patches, PatchCon constructs positive pairs, ordinary negative pairs, and vulnerable--patched hard negative pairs, where patch pairs are emphasized. Each function is encoded by a shared pretrained code encoder, and long functions are processed with chunk-based encoding and max pooling to obtain function-level representations. During training, PatchCon jointly optimizes single-function classification loss and weighted SupCon loss, assigning larger weights to patch-pair hard negatives to better capture subtle differences introduced by fixes. Then the model is evaluated using both classification and pairwise metrics, including Pairwise Accuracy (PACC) and Mean Confidence Gap (MCG), to assess its ability to distinguish vulnerable functions from their patched counterparts.

\subsection{Pair Construction}
\label{subsec:pair_construction}

Based on the vulnerable--patched pairs in the training data, we construct labeled function pairs with explicit semantic relations. Each function pair $(f_q^a, f_q^b)$ is associated with a predefined pair type $t_q$ that characterizes the relation between the two functions. The purpose of this design is to provide both class-level and patch-level supervision for contrastive representation learning. Specifically, we construct four types of function pairs: \textbf{same\_safe} (two non-vulnerable functions, treated as a positive pair), \textbf{same\_vuln} (two vulnerable functions, treated as a positive pair), \textbf{random\_neg} (a randomly sampled vulnerable--non-vulnerable pair, treated as an ordinary negative pair), \textbf{patch} (a vulnerable function and its corresponding patched version, treated as a patch hard negative pair).

The key intuition is that patch pairs provide stronger supervision than ordinary negative pairs. A vulnerable function and its patched counterpart usually share most lexical and syntactic structures, but differ in subtle security-critical semantics. Therefore, distinguishing patch pairs is more challenging and more informative for vulnerability detection than separating randomly sampled vulnerable and non-vulnerable functions. During training, each pair contributes two function instances to a mini-batch. For a mini-batch containing $B$ constructed pairs, we obtain $2B$ function representations. The pair type is retained as metadata and later used by the weighted supervised contrastive objective to distinguish positive pairs, ordinary negative pairs, and patch hard negatives.

\subsection{Pair Input Encoding}
\label{subsec:pair_input_encoding}

Given a function-level code snippet $f$, we first tokenize it into a token sequence:
\begin{equation}
    \mathbf{x} = (x_1, x_2, \ldots, x_n).
\end{equation}
Since real-world functions may exceed the maximum input length of pretrained code models, directly truncating the function may discard vulnerability-relevant statements. To mitigate this issue, we adopt a chunk-based encoding strategy. The token sequence is divided into multiple chunks:
\begin{equation}
    \mathbf{x} \rightarrow \{\mathbf{x}^{(1)}, \mathbf{x}^{(2)}, \ldots, \mathbf{x}^{(m)}\},
\end{equation}
where each chunk is constrained by the maximum input length of the pretrained encoder.

Each chunk is encoded by a shared pretrained code encoder $E(\cdot)$:
\begin{equation}
    \mathbf{c}^{(t)} = E(\mathbf{x}^{(t)}), \quad t = 1,2,\ldots,m,
\end{equation}
where $\mathbf{c}^{(t)} \in \mathbb{R}^{d}$ denotes the representation of the $t$-th chunk. In practice, we use the hidden representation of the special classification token as the chunk-level representation.
To aggregate multiple chunk-level representations into a single function-level representation, we apply element-wise max pooling (MP):
\begin{equation}
    \mathbf{h} = \mathrm{MaxPool}\bigl(\mathbf{c}^{(1)}, \mathbf{c}^{(2)}, \ldots, \mathbf{c}^{(m)}\bigr),
\end{equation}
where $\mathbf{h} \in \mathbb{R}^{d}$ is the final representation of function $f$. MP allows the model to preserve salient vulnerability-related signals that may appear in any chunk of a long function.
For the $q$-th function pair $(f_q^a, f_q^b)$, the two functions are encoded by the same encoder and pooling module with shared parameters:
\begin{equation}
    \mathbf{h}_q^a = \mathrm{Enc}(f_q^a), \quad \mathbf{h}_q^b = \mathrm{Enc}(f_q^b),
\end{equation}
where $\mathrm{Enc}(\cdot)$ denotes the chunk-based code encoder followed by MP. This shared encoding design ensures that both sides of a function pair are represented in the same embedding space.

\subsection{Training Objectives}
\label{subsec:training_objectives}

PatchCon jointly optimizes single-function vulnerability classification and patch-aware SupCon. The classification objective encourages the model to correctly predict whether a function is vulnerable, while the contrastive objective encourages the model to learn a representation space that better captures semantic relations between paired functions.

\subsubsection{Classification Head}

Given a mini-batch of $B$ function pairs, the $q$-th pair is denoted as
$(f_q^a, f_q^b)$.  Although samples are organized as pairs, the classification
head is applied independently to each function:
\begin{equation}
    \hat{\mathbf{y}}_q^a = H_{\mathrm{cls}}(\mathbf{h}_q^a),
    \quad
    \hat{\mathbf{y}}_q^b = H_{\mathrm{cls}}(\mathbf{h}_q^b),
\end{equation}
where $\hat{\mathbf{y}}_q^a, \hat{\mathbf{y}}_q^b \in \mathbb{R}^{2}$ denote
the logits for the non-vulnerable and vulnerable classes. The classification loss averages the cross-entropy losses of the two functions in each pair and then averages over the mini-batch:
\begin{equation}
    \mathcal{L}_{\mathrm{cls}}
    =
    \frac{1}{B}
    \sum_{q=1}^{B}
    \frac{1}{2}
    \left[
    \mathrm{CE}(\hat{\mathbf{y}}_q^a, y_q^a)
    +
    \mathrm{CE}(\hat{\mathbf{y}}_q^b, y_q^b)
    \right],
\end{equation}
where $y_q^a,y_q^b \in \{0,1\}$ are the ground-truth labels.

\subsubsection{Projection Head}
After expanding the $B$ function pairs into $2B$ function instances, the projection head maps each function-level representation into a contrastive embedding space:
\begin{equation}
    \mathbf{z}_i = H_p(\mathbf{h}_i) \in \mathbb{R}^{k},
\end{equation}
where $k$ is the dimension of the contrastive embedding. The projected vector is then $\ell_2$-normalized:
\begin{equation}
    \tilde{\mathbf{z}}_i = \frac{\mathbf{z}_i}{\|\mathbf{z}_i\|_2}.
\end{equation}
The normalized embedding $\tilde{\mathbf{z}}_i$ is used in the supervised contrastive objective.

\subsubsection{Patch-aware Weighted Supervised Contrastive Loss}

Let $\mathcal{V}$ denote the set of $2B$ function instances expanded from the $B$ pairs in a mini-batch. Let $\tilde{\mathbf{z}}_i \in \mathbb{R}^{k}$ denote the normalized embedding of function instance $i \in \mathcal{V}$.
For each anchor $i$, we define its positive set $\mathcal{P}(i)$ according to the constructed positive pair relations:
\begin{equation}
    \mathcal{P}(i) = \{ j \mid (i,j) \text{ forms a same\_safe or same\_vuln pair} \}.
\end{equation}
Pairs of type \textit{random\_neg} and \textit{patch} are treated as negative relations in the contrastive objective. In particular, patch pairs are regarded as hard negatives because they correspond to vulnerable functions and their patched versions. For each anchor $i$ with a non-empty positive set $\mathcal{P}(i)$, the supervised contrastive loss is defined as:
\begin{equation}
    \ell_i =
    - \frac{1}{|\mathcal{P}(i)|}
    \sum_{p \in \mathcal{P}(i)}
    \log
    \frac{
    \exp(\tilde{\mathbf{z}}_i^\top \tilde{\mathbf{z}}_p / \tau)
    }{
    \sum_{a \in \mathcal{V}, a \neq i}
    \exp(\tilde{\mathbf{z}}_i^\top \tilde{\mathbf{z}}_a / \tau)
    }
\end{equation}
where $\tau$ is the temperature hyperparameter. The denominator contains all non-anchor function instances in $\mathcal{V}$, providing uniform normalization over the contrastive candidates.

The supervised contrastive loss is averaged over all anchors with non-empty positive sets:
\begin{equation}
    \mathcal{L}_{\mathrm{pos}} =
    \frac{1}{|\mathcal{I}|}
    \sum_{i \in \mathcal{I}} \ell_i,
\end{equation}
where $\mathcal{I}$ denotes the set of anchor instances with non-empty positive sets. Since only \textit{same\_safe} and \textit{same\_vuln} pairs define positive relations, instances from \textit{patch} pairs do not act as anchors in $\mathcal{L}_{\mathrm{pos}}$. The patch-aware weighting is therefore applied to the explicit negative penalty described below.

In addition to the supervised contrastive formulation above, we introduce an explicit pair-level negative penalty to further separate hard negative pairs. Specifically, for pairs of type \textit{random\_neg} and \textit{patch}, we directly penalize high cosine similarity between the two instances. The pair-level negative loss is defined as:
\begin{equation}
    \mathcal{L}_{\mathrm{neg}}^{w} =
    \frac{1}{|\mathcal{N}|}
    \sum_{(i,j) \in \mathcal{N}}
    w_{ij}
    \cdot
    \mathrm{softplus}
    \left(
        \frac{
        \tilde{\mathbf{z}}_i^{\top} \tilde{\mathbf{z}}_j - m
        }{\tau}
    \right)
\end{equation}
where $\mathcal{N}$ denotes the set of negative pairs (i.e., pairs of type \textit{random\_neg} and \textit{patch}), $m$ is a margin hyperparameter, 
and $w_{ij}$ is the pair-dependent weight, set to $w_p$ for \textit{patch} pairs and $1$ for \textit{random\_neg} pairs, where $w_p>1$ is the hard-negative weight for vulnerable--patched pairs.
The final patch-aware supervised contrastive objective therefore consists of two components:
\begin{equation}
    \mathcal{L}_{\mathrm{supcon}}^{w}
    =
    \mathcal{L}_{\mathrm{pos}}
    +
    \mathcal{L}_{\mathrm{neg}}^{w}
\end{equation}

Overall, the objective combines supervised contrastive learning with an explicit, patch-aware weighted negative repulsion term. Patch pairs are emphasized through larger pair-level weights in the negative penalty, helping the model separate vulnerable functions from their patched counterparts.

\subsubsection{Joint Training Objective}

The overall training objective combines the single-function classification loss and the weighted SupCon loss:
\begin{equation}
\label{ep:final}
    \mathcal{L} =
    \mathcal{L}_{\mathrm{cls}}
    +
    \lambda
    \mathcal{L}_{\mathrm{supcon}}^{w}
\end{equation}
where $\lambda$ controls the contribution of the contrastive objective.

The classification loss encourages the model to predict the vulnerability label of each individual function, while the weighted SupCon loss encourages the model to learn a more discriminative representation space. In particular, the patch-aware weighting strategy increases the contribution of vulnerable--patched hard negatives, enabling the model to focus on subtle security-relevant differences introduced by vulnerability patches. All loss terms are optimized jointly, and gradients are backpropagated through the shared code encoder, MP module, classification head, and projection head.

\section{Experimental Setup}
\label{sec:experimental_setup}

\subsection{Datasets}
\label{subsec:datasets}

We conduct experiments on a realistic vulnerability dataset that contains patch pair information, and further evaluate the trained models on an out-of-distribution (OOD) test set of recently disclosed vulnerabilities.

\textbf{PairVul-Linux}~\citep{du26vulrag} is derived from the benchmark constructed in Vul-RAG. It is built from Linux kernel CVEs and is designed to evaluate whether vulnerability detection models can distinguish vulnerable functions from their corresponding patched versions. It contains 8,628 functions, 4,314 vulnerable--patched pairs, and 2,073 CVEs across 95 different CWEs. Each instance provides a vulnerable function and a patched function, together with vulnerability-related metadata such as CVE ID, CWE ID, and patched code. In this work, PairVul-Linux serves as the primary benchmark because its paired structure naturally supports both function-level vulnerability classification and pairwise discrimination. 

\textbf{VentiVul}~\citep{lu2025from} is a temporally separated OOD test set of recently disclosed Linux and Chromium vulnerabilities from our previous work. We use its Linux kernel subset, which follows the construction procedure of PairVul~\citep{du26vulrag} and contains the functions modified by the fixing commits of Linux kernel CVEs published between May 1 and May 29, 2025, as vulnerable--patched pairs. This subset contains 116 pairs from 100 CVEs across 20 CWEs. No CVE in VentiVul appears in PairVul-Linux, whose most recent CVE (CVE-2024-0193) was published in January 2024. To further avoid code-level overlap, we remove four pairs whose two versions are identical after whitespace normalization and one pair whose vulnerable function also appears in PairVul-Linux, resulting in 111 pairs from 96 CVEs, which we use only for testing and refer to as VentiVul in the following.

\subsection{Baseline Models}
\label{subsec:baselines}

We compare PatchCon with three pretrained code-model baselines and several representative vulnerability detection methods that are closely related to CL or patch-pair-based vulnerability analysis.

\textbf{CodeBERT}~\citep{codebert20}, \textbf{GraphCodeBERT}~\citep{guo2021graphcb}, and \textbf{UniXcoder}~\citep{unixcoder22} are used as pretrained code model baselines to study the effect of different encoders on vulnerability detection. CodeBERT is a bimodal Transformer pretrained on code and natural language, and serves as our function-level classification baseline and the default backbone of PatchCon. GraphCodeBERT enhances code representation with data-flow information and structure-aware pretraining, while UniXcoder supports cross-modal code understanding and generation by leveraging multimodal signals such as code comments.

\textbf{ReVeal}~\citep{chakraborty22deep} is a Graph Neural Network (GNN)-based approach for vulnerability detection. It represents source code with graph-based features and introduces a contrastive learning mechanism through triplet loss, making it an early attempt to incorporate contrastive learning into vulnerability detection. We include ReVeal as a representative graph-based and contrastive-learning-based vulnerability detection baseline.

\textbf{ContraBERT}~\citep{contrabert23} introduces InfoNCE-based CL to improve the semantic structure of code embeddings. It provides both CodeBERT- and GraphCodeBERT-based variants; in this work, we adopt the CodeBERT-based version to keep the backbone consistent with our main experimental setting.

\textbf{SCL-CVD}~\citep{wang24sclcvd} is a supervised contrastive learning-based vulnerability detection method. It fine-tunes a pretrained code model with supervised contrastive learning and uses a multilayer perceptron (MLP) classifier for vulnerability prediction. However, unlike PatchCon, it does not explicitly model vulnerable--patched pairs as weighted hard negatives. As its public implementation does not include the GraphCodeBERT-based module, we replicate its CodeBERT version.

\textbf{DELTA}~\citep{tian26when} is a contrastive learning method for optimization-resilient binary vulnerability detection. It constructs contrastive samples from vulnerable and patched function pairs, and fine-tunes CodeBERT with contrastive learning. As DELTA is originally designed for binary code, we reproduce its source-code training version for our setting.

\textbf{Vul-RAG}~\citep{du26vulrag} is a retrieval-augmented LLM framework for vulnerability detection. It incorporates vulnerability knowledge distilled from historical vulnerabilities and fixes to help LLMs distinguish vulnerable code from similar patched code. Vul-RAG can be instantiated with different LLM backbones; in this work, we report the best-performing version, Vul-RAG with GPT-4o, which achieves the highest pairwise accuracy among the four LLMs evaluated in the original paper and whose per-sample outputs are publicly released. Unless otherwise specified, ``Vul-RAG'' in the following sections refers to this GPT-4o-based version.

\subsection{Implementation Details}
\label{subsec:implementation}

We implement PatchCon with PyTorch 2.3.1 and HuggingFace Transformers 4.37.2 under Python 3.12.4 and CUDA 11.8.0. All experiments are conducted on a multi-GPU server equipped with NVIDIA H100 GPUs. Unless otherwise stated, CodeBERT is the default encoder. We use AdamW with learning rate $2\times10^{-5}$, batch size 16, SupCon temperature $\tau=0.07$, chunk length 512, and MP size 2048. Hyperparameters and model checkpoints are selected according to the performance on the validation set, and the final objective in Eq.~(\ref{ep:final}) uses $\lambda=0.01$ and patch-pair weight $w_p=2$. Training PatchCon takes about 53 minutes per epoch on a single H100 GPU (at most 10 epochs with early stopping), and inference takes about 23 ms per function. PairVul-Linux is randomly split into training, validation, and test sets with an 8:1:1 ratio at the pair level, so that the two functions of each patch pair always fall into the same partition and the class balance is preserved. As in the original PairVul benchmark~\citep{du26vulrag}, the split is performed at the pair level rather than grouped by CVE; we discuss the effect of CVE overlap between partitions in Section~\ref{subsec:cve_overlap}.

\subsection{Evaluation Metrics}
\label{subsec:metrics}

We use standard binary classification metrics, including Accuracy (ACC), Precision, Area Under the Precision-Recall Curve ($\text{AUC}_{\text{PRC}}$) and Matthews Correlation Coefficient (MCC). Since our evaluation sets consist of vulnerable--patched pairs and are therefore class-balanced, Accuracy and MCC are reported as primary threshold-based metrics. Precision is included to characterize prediction exactness, while $\text{AUC}_{\text{PRC}}$, computed from the predicted vulnerability probabilities, summarizes the precision--recall trade-off across thresholds. Recall and F1 are reported as supplementary metrics rather than primary metrics because they may obscure asymmetric behavior when a model over-predicts the vulnerable class. 
Let true positives ($TP$), true negatives ($TN$), false positives ($FP$), and false negatives ($FN$) denote the elements of the confusion matrix. The $\text{AUC}_{\text{PRC}}$ is computed as:
\begin{equation}
    \text{AUC}_{\text{PRC}} = \int_{0}^{1} P(R) \, dR,
\end{equation}
where $P(R)$ denotes Precision as a function of Recall $R$. MCC is computed as:
\begin{equation}
    \mathrm{MCC} =
    \frac{TP \cdot TN - FP \cdot FN}
    {\sqrt{(TP + FP)(TP + FN)(TN + FP)(TN + FN)}}.
\end{equation}

To evaluate vulnerable--patched discrimination, we use two pairwise metrics. PACC corresponds to the pair-wise correct prediction metric of PrimeVul~\citep{primevul25} and the pairwise accuracy of Vul-RAG~\citep{du26vulrag}, while MCG complements it from a confidence-based perspective. Given a patch pair $(f_v, f_p)$, where $f_v$ denotes the vulnerable function and $f_p$ denotes the patched function, let $s(f)$ denote the predicted vulnerability confidence of function $f$, and let $\hat{y}(f)$ denote its predicted label.

\textbf{Pairwise Accuracy (PACC)} measures the proportion of patch pairs that are correctly distinguished at the prediction-label level. A patch pair is considered successful only when the vulnerable function is predicted as vulnerable and its patched counterpart is predicted as safe:
\begin{equation}
    \mathrm{PACC} =
    \frac{1}{|\mathcal{P}_{\mathrm{patch}}|}
    \sum_{(f_v,f_p) \in \mathcal{P}_{\mathrm{patch}}}
    \mathbb{I}\bigl(\hat{y}(f_v)=1 \land \hat{y}(f_p)=0\bigr),
\end{equation}
where $\mathcal{P}_{\mathrm{patch}}$ denotes the set of vulnerable--patched function pairs. Following PrimeVul~\citep{primevul25}, we further divide the pairs that are not correctly distinguished into three failure modes: both functions predicted as vulnerable (P-V), both predicted as safe (P-B), and reversed predictions (P-R).

\textbf{Mean Confidence Gap (MCG)} further measures the average difference in vulnerability confidence between vulnerable functions and their patched counterparts:
\begin{equation}
    \mathrm{MCG} =
    \frac{1}{|\mathcal{P}_{\mathrm{patch}}|}
    \sum_{(f_v,f_p) \in \mathcal{P}_{\mathrm{patch}}}
    \bigl(s(f_v) - s(f_p)\bigr).
\end{equation}
A higher MCG indicates that the model assigns a larger vulnerability confidence margin to vulnerable functions over their patched counterparts. Together, PACC and MCG evaluate pairwise vulnerable--patched discrimination from complementary perspectives.

\section{Evaluation}
\label{sec:evaluation}
Our evaluation is designed to answer the following research questions.
\begin{itemize}
    \item \textbf{RQ1.} How effective is PatchCon compared with state-of-the-art vulnerability detection approaches?
    \item \textbf{RQ2.} How do PatchCon's components and backbone choices affect performance?
    \item \textbf{RQ3.} How do key hyperparameters affect PatchCon's performance?
    \item \textbf{RQ4.} How does PatchCon perform across CWE types, function lengths, and patch sizes?
    \item \textbf{RQ5.} Are PatchCon's improvements statistically significant?
    \item \textbf{RQ6.} Does PatchCon's vulnerable--patched discrimination generalize to vulnerabilities disclosed after its training data?
\end{itemize}
Section~\ref{sec:discussion} further discusses the robustness of the results to CVE overlap and representative cases.

\subsection{Overall Performance (RQ1)}

Table~\ref{tab:overall_performance} compares PatchCon with representative baselines on PairVul-Linux, including pretrained code-model baselines, CL-based vulnerability detection methods, and a prompt-based LLM method.
As the Vul-RAG configuration uses the closed-source GPT-4o model, we do not rerun it within our experimental pipeline. Instead, we compute its metrics from the per-sample outputs released by its authors, which cover 585 of the 586 pairs in its own test split of PairVul. Since Vul-RAG outputs only binary decisions, $\text{AUC}_{\text{PRC}}$ and MCG cannot be computed and are marked as ``--'', and undetermined answers, which occur in 19 pairs, are counted as incorrect predictions. The recomputed PACC (0.2923) differs from the 0.32 reported in the original paper; we use the recomputed values so that all metrics of Vul-RAG are derived from the same predictions. Since these results were obtained on a different test split of PairVul, they are not directly comparable with ours. Overall, PatchCon achieves the strongest performance among the evaluated models, especially on MCC and pairwise metrics. Compared with the CodeBERT (CB) baseline, PatchCon improves ACC from 0.5174 to 0.6030 and MCC from 0.0596 to 0.2060. The gains are more pronounced on pairwise metrics, where PACC increases from 0.0486 to 0.3264 and MCG increases from 0.0035 to 0.1838. Similar trends are observed for GraphCodeBERT (GraphCB) and UniXcoder (UX), whose PACC and MCG remain close to zero.

\begin{table}[ht]
\centering
\caption{Overall performance comparison with representative baselines on the PairVul-Linux test set.
Pair: uses vulnerable--patched function pairs (for Vul-RAG, to build its knowledge base);
PW: patch-pair weighting; InfoNCE, Triplet, NT-Xent, and SupCon also represent contrastive loss types.}
\label{tab:overall_performance}
\begin{adjustbox}{max width=\textwidth}
\begin{tabular}{lllcccccccc}
\toprule
\textbf{Approach} &
\textbf{Backbone} &
\textbf{Contrastive type} &
\textbf{Pair} &
\textbf{PW} &
\textbf{ACC} &
\textbf{Precision} &
\textbf{MCC} &
\textbf{$\text{AUC}_{\text{PRC}}$} &
\textbf{PACC} &
\textbf{MCG} \\
\midrule
UniXcoder      & UX  & $\times$  & $\times$  & $\times$  & 0.5023 & 0.5012 & 0.0482 & 0.5355 & 0.0046 & 0.0024 \\
GraphCodeBERT  & GraphCB & $\times$  & $\times$  & $\times$  & 0.5069 & 0.5035 & 0.0486 & 0.5363 & 0.0185 & 0.0076 \\
CodeBERT       & CB  & $\times$  & $\times$  & $\times$  & 0.5174 & 0.5096 & 0.0596 & 0.5329 & 0.0486 & 0.0035 \\
ContraBERT     & CB	 & InfoNCE & $\times$  & $\times$  & 0.5150 & 0.5096 & 0.0364 & 0.5312 & 0.0671 & 0.0026 \\
ReVeal$^{\mathrm{a}}$ & GNN & Triplet & $\times$  & $\times$  & 0.5024 & 0.5017 & 0.0052 & 0.5022 & 0.0481 & 0.0000 \\
SCL-CVD        & CB  & SupCon  & $\times$ & $\times$  & 0.5544 & 0.5540 &	0.1088 & 0.5561 & 0.1505 & 0.0271 \\
DELTA          & CB  & SimCLR NT-Xent & $\checkmark$ & $\times$  & 0.5729 & 0.5609 & 0.1487 & 0.5786 & 0.2060 & 0.0855\\
Vul-RAG$^{\mathrm{b}}$ & GPT-4o & $\times$  & $\checkmark$ & $\times$  & 0.5974 & 0.6004 & 0.1950 & --  & 0.2923 & --  \\
\midrule
\textbf{PatchCon}   & CB  & Weighted SupCon & $\checkmark$ & $\checkmark$  & \textbf{0.6030} & \textbf{0.6018} & \textbf{0.2060} & \textbf{0.6239} & \textbf{0.3264} & \textbf{0.1838} \\
\bottomrule
\end{tabular}
\end{adjustbox}
\tabnotes{$^{\mathrm{a}}$Graph construction failed for 16 of the 432 test pairs, so ReVeal is evaluated on the remaining 416 pairs.\\
$^{\mathrm{b}}$Computed from the per-sample outputs released by the authors of Vul-RAG for 585 of the 586 pairs in its own test split, with undetermined answers counted as incorrect; $\text{AUC}_{\text{PRC}}$ and MCG cannot be computed from its binary outputs.}
\end{table}

\begin{figure}
    \centering
    \includegraphics[width=1\linewidth]{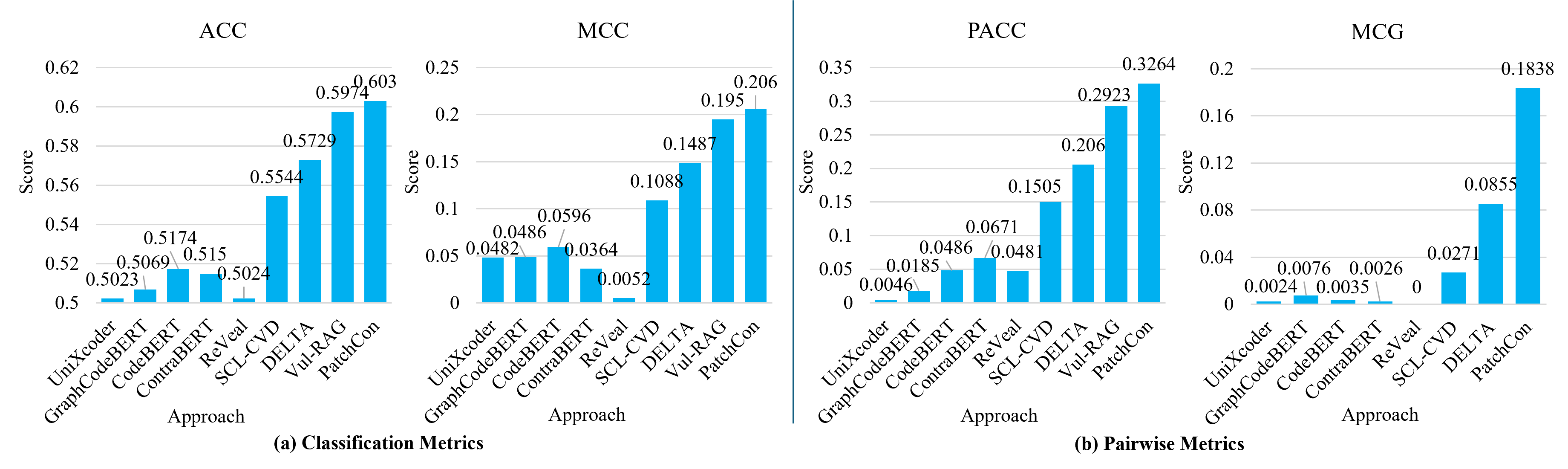}
    \caption{Overall performance comparison of different models.}
    \label{fig:overall_perf}
\end{figure}

Compared with CL-based baselines, PatchCon also shows clear advantages. ContraBERT, which introduces InfoNCE-based contrastive learning, brings only marginal changes over CodeBERT across all metrics, suggesting that general contrastive pretraining alone may not be sufficient for fine-grained vulnerable--patched discrimination. ReVeal uses triplet loss but is not designed to explicitly model patch pairs, which may explain its lower pairwise performance in this setting. SCL-CVD and DELTA improve over ordinary classification by using supervised CL and pair-based CL, respectively, reaching PACC scores of 0.1505 and 0.2060, respectively, which PatchCon further improves to 0.3264. Vul-RAG obtains a competitive PACC of 0.2923 under a prompt-based LLM setting; however, it relies on closed-source LLMs, and its result is obtained on a different test split. PatchCon achieves a comparable PACC (0.3264) while also yielding a substantially positive MCG, which cannot be computed from the binary outputs of Vul-RAG. To compare the two methods on identical inputs, we further identify the 59 pairs that appear in both our test set and the test split of Vul-RAG. On these pairs, PatchCon achieves a PACC of 0.3729 and an MCC of 0.2888, compared with 0.2881 and 0.2381 for Vul-RAG, although this subset is small. In addition, PatchCon is lightweight: it builds on a 125M-parameter encoder and requires about 23 ms of inference per function on a single GPU, whereas Vul-RAG issues on average 2.3 GPT-4o calls per function at detection time (about 2.0k input and 2.2k output tokens per function according to its released outputs), in addition to the LLM calls used for knowledge extraction.
Figure~\ref{fig:overall_perf} visually summarizes the same trend across ACC, MCC, PACC, and MCG.

Table~\ref{tab:diagnostics} further reports Recall, F1, and the distribution of pairwise failure modes. The pretrained encoders obtain high Recall and F1 mainly because they predict almost all functions as vulnerable, so most of their pairs fall into P-V (e.g., 0.9954 for UniXcoder and 0.8750 for CodeBERT). PatchCon distributes its remaining errors more evenly between P-V (0.2824) and P-B (0.2708). Its proportion of reversed predictions (0.1204) is the highest among the models, as models that predict almost all functions as vulnerable rarely produce reversed pairs; nevertheless, most of its failures stem from assigning the same label to both versions of a function rather than inverting them.

\begin{table}[htbp]
\centering
\caption{Supplementary metrics on the PairVul-Linux test set. P-C: both functions of a patch pair are correctly classified (equal to PACC); P-V: both are predicted as vulnerable; P-B: both are predicted as safe; P-R: the predictions are reversed.}
\label{tab:diagnostics}
\begin{adjustbox}{max width=\textwidth}
\begin{tabular}{lcccccc}
\toprule
\textbf{Approach} & \textbf{Recall} & \textbf{F1} & \textbf{P-C} & \textbf{P-V} & \textbf{P-B} & \textbf{P-R} \\
\midrule
UniXcoder          & 1.0000 & 0.6677 & 0.0046 & 0.9954 & 0.0000 & 0.0000 \\
GraphCodeBERT      & 0.9861 & 0.6667 & 0.0185 & 0.9676 & 0.0093 & 0.0046 \\
CodeBERT           & 0.9236 & 0.6568 & 0.0486 & 0.8750 & 0.0625 & 0.0139 \\
ContraBERT         & 0.7963 & 0.6215 & 0.0671 & 0.7292 & 0.1667 & 0.0370 \\
ReVeal$^{\mathrm{a}}$ & 0.6899 & 0.5810 & 0.0481 & 0.6418 & 0.2668 & 0.0433 \\
SCL-CVD            & 0.5579 & 0.5559 & 0.1505 & 0.4074 & 0.4005 & 0.0417 \\
DELTA              & 0.6713 & 0.6112 & 0.2060 & 0.4653 & 0.2685 & 0.0602 \\
Vul-RAG$^{\mathrm{b}}$ & 0.5829 & 0.5915 & 0.2923 & 0.2906 & 0.3197 & 0.0974 \\
\midrule
\textbf{PatchCon}  & 0.6088 & 0.6053 & \textbf{0.3264} & 0.2824 & 0.2708 & 0.1204 \\
\bottomrule
\end{tabular}
\end{adjustbox}
\tabnotes{$^{\mathrm{a}}$Evaluated on the 416 test pairs available for ReVeal.\\
$^{\mathrm{b}}$Computed from the released outputs of Vul-RAG on its own test split (585 pairs; see Table~\ref{tab:overall_performance}).}
\end{table}

\begin{answerbox}
\textbf{Answer to RQ1.} PatchCon achieves the strongest overall performance on PairVul-Linux, especially on MCC, PACC, and MCG, showing more reliable vulnerable--patched discrimination than the evaluated baselines.
\end{answerbox}

\subsection{Ablation Study (RQ2)}
For all ablation settings, we use the same training/validation/test splits, optimizer, batch size, learning rate, and random seeds.
Each ablation removes or modifies only the corresponding component of the full model.

\subsubsection{PatchCon Ablation}
Table~\ref{tab:ablation} reports the ablation results of different variants of the proposed framework. Overall, the full PatchCon model achieves the best performance on all metrics except $\text{AUC}_{\text{PRC}}$, where the unweighted SupCon variant is slightly higher. Compared with the CB baseline, PatchCon improves performance on both classification and pairwise metrics. These results indicate that the proposed framework not only improves vulnerability classification, but also substantially strengthens the model's ability to distinguish vulnerable functions from their patched counterparts. The comparison between CB and PairCodeBERT (a pair-input variant introduced in this work by adapting CodeBERT to constructed function pairs) shows that pair-based input construction alone brings no consistent gains in standard metrics, but improves PACC from 0.0486 to 0.0903 and MCG from 0.0035 to 0.0176, suggesting that paired inputs provide useful discriminative signals.
Adding MP further improves both classification and pairwise metrics, confirming the importance of preserving vulnerability-relevant signals from long functions.
The degradation observed in the ``w/o MP'' variant further supports this conclusion, where removing MP reduces ACC to 0.5463 and PACC to 0.1551. However, the improvements of PatchCon do not stem from MP alone. MP only changes the representations of functions that exceed a single 512-token chunk, and on the 224 test pairs (51.9\%) whose two versions both fit into a single chunk, PatchCon still achieves a PACC of 0.3527, compared with 0.2723 for DELTA and 0.2009 for SCL-CVD (see also Section~\ref{subsec:cwe_specific_analysis}).

\begin{table}[htbp]
\centering
\caption{Performance comparison of different variants of PatchCon. 
MP: max pooling; Patch: patch-pair input;
$\checkmark$ indicates that the component is used, $\times$ indicates that it is not used, and
$\circ$ indicates that only a subset of pair types is retained.
}
\label{tab:ablation}
\begin{adjustbox}{max width=\textwidth}
\begin{tabular}{lccccccccccc}
\toprule
\textbf{Approach} & \textbf{MP} & \textbf{Pair} & \textbf{Patch} & \textbf{SupCon} & \textbf{Weight} & \textbf{ACC} & \textbf{Precision} & \textbf{MCC} & \textbf{$\text{AUC}_{\text{PRC}}$} & \textbf{PACC} & \textbf{MCG} \\
\midrule
CodeBERT & $\times$  & $\times$ & $\times$ & $\times$ & -- & 0.5174 & 0.5096 & 0.0596 & 0.5329 & 0.0486 & 0.0035 \\
PairCodeBERT & $\times$ & $\checkmark$ & $\checkmark$ & $\times$ & -- & 0.5197 & 0.5206 & 0.0394 & 0.5382 & 0.0903 & 0.0176 \\
PairCodeBERT + MP & $\checkmark$ & $\checkmark$ & $\checkmark$ & $\times$ & -- & 0.5729 & 0.5897 & 0.1485 & 0.5981 & 0.2384 & 0.1310 \\
PairCodeBERT + SupCon & $\checkmark$ & $\checkmark$ & $\checkmark$ & $\checkmark$ & 1 & 0.5706 & 0.5840 & 0.1430 & \textbf{0.6367} & 0.2708 & 0.1499 \\
\midrule
w/o MP & $\times$ & $\checkmark$ & $\checkmark$ & $\checkmark$ & 2 & 0.5463 & 0.5629 & 0.0960 & 0.5614 & 0.1551 & 0.0561 \\
w/o patch pairs & $\checkmark$ & $\checkmark$ & $\times$ & $\checkmark$ & -- & 0.5104 & 0.5102 & 0.0208 & 0.5298 & 0.0718 & 0.0039 \\
w/o random\_neg pairs & $\checkmark$ & $\circ$ & $\checkmark$ & $\checkmark$ & 2 & 0.5428 & 0.5437 & 0.0857 & 0.5498 & 0.1435 & 0.0187 \\
w/o positive pairs & $\checkmark$ & $\circ$ & $\checkmark$ & $\checkmark$ & 2 & 0.5752 & 0.5900 & 0.1525 & 0.6128 & 0.2755 & 0.1479 \\
\midrule
PatchCon (full model) & $\checkmark$ & $\checkmark$ & $\checkmark$ & $\checkmark$ & 2 & \textbf{0.6030} & \textbf{0.6018} & \textbf{0.2060} & 0.6239 & \textbf{0.3264} & \textbf{0.1838} \\
\bottomrule
\end{tabular}
\end{adjustbox}
\end{table}

The results further validate the effectiveness of SupCon and patch-aware weighting. Compared with PairCodeBERT with MP, the unweighted SupCon variant achieves similar ACC and MCC but improves PACC from 0.2384 to 0.2708 and MCG from 0.1310 to 0.1499, suggesting that SupCon helps learn representations that better separate vulnerable and patched functions. Applying patch-aware weighting further improves MCC from 0.1430 to 0.2060 and PACC from 0.2708 to 0.3264, showing that patch pairs provide informative hard negatives due to their lexical and syntactic similarity but distinct security semantics. Pair-type ablations confirm that patch pairs are crucial for learning security-sensitive distinctions: removing them causes the largest degradation, reducing PACC to 0.0718 and MCG to 0.0039, close to the CodeBERT baseline. Removing random negatives also reduces PACC to 0.1435, indicating that random negatives are important for stabilizing the general vulnerable/non-vulnerable boundary. 
Although removing positive pairs still yields reasonable results, the full model remains consistently better, indicating that positive pairs also contribute to intra-class compactness. Overall, max pooling, pair construction, SupCon, and patch-aware weighting are complementary, and their combination achieves the strongest classification and pairwise discrimination performance.

\subsubsection{PatchCon Variants with Different Backbone}

Table~\ref{tab:backbone} compares PatchCon with different pretrained code backbones under a fixed contrastive setting ($w_p=3$, $\lambda=0.01$), i.e., before the hard-negative weight was set to its final value $w_p=2$. Overall, CodeBERT performs best in this setting. Without SupCon, the CodeBERT-based PatchCon already outperforms the other backbones in the same setting, achieving 0.5729 ACC, 0.5897 Precision, 0.1485 MCC, 0.5981 $\text{AUC}_{\text{PRC}}$, 0.2384 PACC, and 0.1310 MCG. With SupCon, it further achieves the best results across all metrics, suggesting that CodeBERT provides stronger initial representations for the patch-aware contrastive framework. Across all three backbones, SupCon consistently improves performance, especially on PACC and MCG, suggesting the backbone-agnostic effectiveness of the proposed patch-aware contrastive strategy, while final performance still depends on encoder quality.

\begin{table}[ht]
\centering
\caption{Performance of PatchCon with different pretrained code backbones with MP size 2048 and SupCon loss $\lambda=0.01$, and patch-pair weight $w_p=3$.}
\label{tab:backbone}
\begin{adjustbox}{max width=\textwidth}
\begin{tabular}{llcccccc}
\toprule
\textbf{Backbone} & \textbf{SupCon} & \textbf{ACC} & \textbf{Precision} & \textbf{MCC} & \textbf{$\text{AUC}_{\text{PRC}}$} & \textbf{PACC} & \textbf{MCG} \\
\midrule
\multirow{2}{*}{CodeBERT}      & $\times$     & 0.5729 & 0.5897 & 0.1485 & 0.5981 & 0.2384 & 0.1310 \\
                               & $\checkmark$ & \textbf{0.5961} & \textbf{0.6040} & \textbf{0.1927} & \textbf{0.6208} & \textbf{0.3218} & \textbf{0.1715} \\
\midrule
\multirow{2}{*}{GraphCodeBERT} & $\times$     & 0.5058 & 0.5034 & 0.0158 & 0.5275 & 0.0671 & 0.0050 \\
                               & $\checkmark$ & \textbf{0.5613} & \textbf{0.5556} & \textbf{0.1234} & \textbf{0.5683} & \textbf{0.2870} & \textbf{0.1179} \\
\midrule
\multirow{2}{*}{UniXcoder}     & $\times$     & 0.4977 & 0.4977 & -0.0046 & 0.5384 & 0.0856 & 0.0086 \\
                               & $\checkmark$ & \textbf{0.5637} & \textbf{0.5610} & \textbf{0.1274} & \textbf{0.6006} & \textbf{0.2593} & \textbf{0.1221} \\
\bottomrule
\end{tabular}
\end{adjustbox}
\end{table}

\begin{answerbox}
\textbf{Answer to RQ2.} The ablation and backbone results show that MP, pair construction, SupCon, and patch-aware weighting provide complementary benefits, and that the proposed contrastive strategy improves performance across different pretrained code backbones.
\end{answerbox}

\subsection{Hyperparameter Sensitivity (RQ3)}
\label{subsec:parameter_analysis}
This RQ examines how sensitive PatchCon is to its three main hyperparameters: the SupCon loss weight $\lambda$, the hard negative weight $w_p$, and the max pooling size. Starting from the default configuration described in Section~\ref{subsec:implementation}, we vary one hyperparameter at a time and report the test-set results in Tables~\ref{tab:supcon_lambda}--\ref{tab:pooling_size}. Because the hyperparameters were tuned sequentially, $\lambda$ is varied with $w_p=3$, $w_p$ with $\lambda=0.01$, and the pooling size with $\lambda=0.01$ and $w_p=2$.

\subsubsection{Effect of SupCon Loss}
\label{subsubsec:supcon_lambda}
Table~\ref{tab:supcon_lambda} shows that PatchCon is robust to $\lambda$ within a moderate range. For $\lambda$ between 0.005 and 0.05, PACC ranges from 0.2639 to 0.3218 and MCC from 0.1554 to 0.1984, all above the variant trained without SupCon (PACC 0.2384 and MCC 0.1485; Table~\ref{tab:ablation}). Performance drops sharply at $\lambda=0.1$ (PACC 0.1458 and MCC 0.1016), suggesting that an overly strong contrastive term weakens the classification objective.

\begin{table}[htbp]
\centering
\caption{Sensitivity of PatchCon to the SupCon loss weight $\lambda$ on the test set, with MP size 2048 and patch-pair weight $w_p=3$.}
\label{tab:supcon_lambda}
\begin{adjustbox}{max width=\textwidth}
\begin{tabular}{ccccccc}
\toprule
\textbf{$\lambda$} &
\textbf{ACC} &
\textbf{Precision} &
\textbf{MCC} &
\textbf{$\text{AUC}_{\text{PRC}}$} &
\textbf{PACC} &
\textbf{MCG} \\
\midrule
0.005 & 0.5775 & 0.5730 & 0.1554 & 0.6174 & 0.2639 & 0.1501 \\
0.01 (default) & \textbf{0.5961} & \textbf{0.6040} & 0.1927 & 0.6208 & \textbf{0.3218} & 0.1715 \\
0.05  & 0.5938 & 0.5707 & \textbf{0.1984} & \textbf{0.6235} & 0.2917 & \textbf{0.1723} \\
0.1   & 0.5463 & 0.5787 & 0.1016 & 0.5737 & 0.1458 & 0.0478 \\
\bottomrule
\end{tabular}
\end{adjustbox}
\end{table}

\subsubsection{Effect of Hard Negative Weight}
Table~\ref{tab:patch_weight} shows that weighting patch pairs more heavily than ordinary negatives is beneficial within a limited range. With $w_p=2$ and $w_p=3$, PACC reaches 0.3264 and 0.3218 and MCC 0.2060 and 0.1927, compared with 0.2708 and 0.1430 when patch pairs are treated as ordinary negatives ($w_p=1$), although $w_p=1$ obtains the highest $\text{AUC}_{\text{PRC}}$. A smaller increase ($w_p=1.5$) brings no clear gain, whereas $w_p=4$ reduces PACC to 0.1736, indicating that excessive emphasis on patch pairs disrupts the balance between hard negative separation and classification.

\begin{table}[htbp]
\centering
\caption{Sensitivity of PatchCon to the patch-pair hard negative weight $w_p$ on the test set, with MP size 2048 and $\lambda=0.01$.}
\label{tab:patch_weight}
\begin{adjustbox}{max width=\textwidth}
\begin{tabular}{ccccccc}
\toprule
\textbf{$w_p$} &
\textbf{ACC} &
\textbf{Precision} &
\textbf{MCC} &
\textbf{$\text{AUC}_{\text{PRC}}$} &
\textbf{PACC} &
\textbf{MCG} \\
\midrule
1   & 0.5706 & 0.5840 & 0.1430 & \textbf{0.6367} & 0.2708 & 0.1499 \\
1.5 & 0.5764 & 0.5965 & 0.1562 & 0.6159 & 0.2616 & 0.1409 \\
2 (default) & \textbf{0.6030} & 0.6018 & \textbf{0.2060} & 0.6239 & \textbf{0.3264} & \textbf{0.1838} \\
3  & 0.5961 & \textbf{0.6040} & 0.1927 & 0.6208 & 0.3218 & 0.1715 \\
4   & 0.5451 & 0.5575 & 0.0924 & 0.5792 & 0.1736 & 0.0581 \\
\bottomrule
\end{tabular}
\end{adjustbox}
\end{table}

\subsubsection{Effect of Pooling Size}
\label{subsubsec:pooling_size}

Table~\ref{tab:pooling_size} shows that the pooling size has the strongest effect. PACC increases from 0.1551 with a single 512-token chunk to 0.2755 with 1,536 tokens and 0.3264 with 2,048 tokens, the largest size we evaluate, whereas 1,024 tokens performs worse than 512 tokens (PACC 0.0880), possibly because additional code context introduces spurious chunk-level activations into the max-pooled representation. This indicates that sufficient coverage of long functions is important for preserving vulnerability-relevant signals, consistent with the analysis by function length in RQ4.

\begin{table}[htbp]
\centering
\caption{Sensitivity of PatchCon to the max pooling size on the test set, with $\lambda=0.01$ and patch-pair weight $w_p=2$.}
\label{tab:pooling_size}
\begin{adjustbox}{max width=\textwidth}
\begin{tabular}{ccccccc}
\toprule
\textbf{Size} &
\textbf{ACC} &
\textbf{Precision} &
\textbf{MCC} &
\textbf{$\text{AUC}_{\text{PRC}}$} &
\textbf{PACC} &
\textbf{MCG} \\
\midrule
512  & 0.5463 & 0.5629 & 0.0960 & 0.5614 & 0.1551 & 0.0561 \\
1024 & 0.5220 & 0.5212 & 0.0440 & 0.5293 & 0.0880 & 0.0054 \\
1536 & 0.5961 & 0.5819 & 0.1951 & 0.6124 & 0.2755 & 0.1560 \\
2048 (default) & \textbf{0.6030} & \textbf{0.6018} & \textbf{0.2060} & \textbf{0.6239} & \textbf{0.3264} & \textbf{0.1838} \\
\bottomrule
\end{tabular}
\end{adjustbox}
\end{table}

\begin{answerbox}
\textbf{Answer to RQ3.} PatchCon is robust to the SupCon loss weight between 0.005 and 0.05 and to the hard negative weight between 2 and 3, but degrades when either weight is too large. The pooling size has the strongest effect, and covering up to 2,048 tokens gives the best results among the evaluated sizes.
\end{answerbox}

\subsection{Analysis by CWE Type, Function Length, and Patch Size (RQ4)}
\label{subsec:cwe_specific_analysis}

\begin{figure}
    \centering
    \includegraphics[width=1\linewidth]{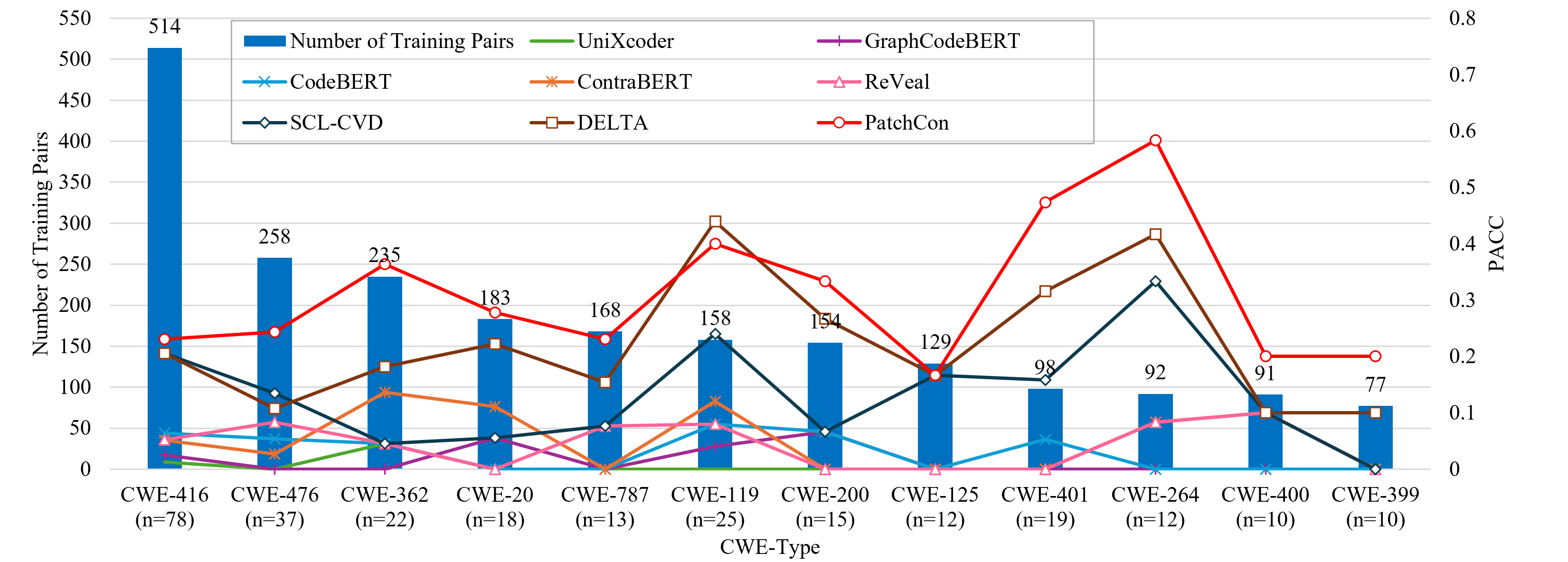}
    \caption{CWE-specific PACC comparison across the 12 most frequent CWE types in PairVul-Linux. Bars denote the number of training pairs for each CWE type, lines denote the PACC achieved by each model on the test set, and $n$ denotes the number of test pairs.}
\label{fig:cwe_specific_pacc}
\end{figure}

To further examine whether the improvement of PatchCon is concentrated in specific vulnerability types or generalizes across different CWEs, we conduct a CWE-specific analysis based on patch-pair-level PACC. Fig.~\ref{fig:cwe_specific_pacc} reports the results on the 12 most frequent CWE types in PairVul-Linux. Overall, PatchCon achieves the best or tied-best PACC on 11 out of the 12 CWE types, indicating that the proposed patch-aware SupCon strategy improves vulnerable--patched discrimination across most vulnerability categories. For the four CWE types with the most training pairs (CWE-416, CWE-476, CWE-362, and CWE-20), PatchCon obtains PACC scores of 0.2308, 0.2432, 0.3636, and 0.2778, respectively, outperforming all compared baselines. These results suggest that explicitly modeling vulnerable--patched pairs as hard negatives helps the model capture subtle security-relevant differences across multiple CWE categories.
The relative improvement of PatchCon still varies across CWE types. For CWE-416, which has the most training pairs (514) and test pairs (78), SCL-CVD and DELTA already outperform vanilla pretrained encoders, but PatchCon further improves PACC to 0.2308. For several CWE types where most baselines obtain very low or even zero PACC, PatchCon achieves clear gains on different CWE types. The only exception is CWE-119 (25 test pairs), where DELTA achieves 0.4400 while PatchCon obtains 0.4000. Moreover, the per-CWE PACC of PatchCon is not associated with the amount of training data (Spearman's $\rho=0.09$ across the 12 CWE types): CWE-416 has the most training pairs but a relatively low PACC, whereas CWE-264, with only 92 training pairs, reaches 0.5833. This indicates that the difficulty of a CWE type depends more on the characteristics of its vulnerabilities and fixes than on the number of training pairs. These findings suggest that the improvement of PatchCon is not limited to aggregate performance but extends to most vulnerability categories.

\begin{figure}[htbp]
\centering
\includegraphics[width=0.75\linewidth]{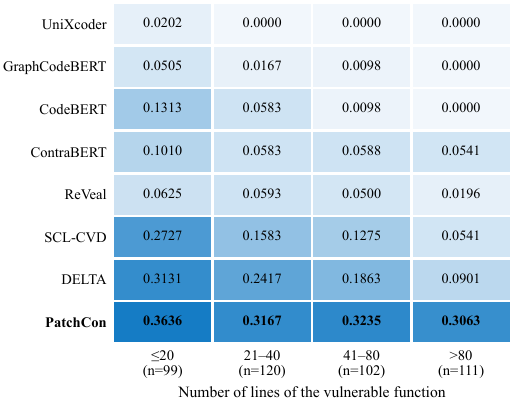}
\caption{PACC on the PairVul-Linux test set grouped by the number of lines of the vulnerable function; $n$ denotes the number of test pairs in each group. ReVeal is evaluated on the pairs for which graphs can be constructed.}
\label{fig:pacc_length}
\end{figure}

We further group the test pairs into four bins of similar size according to the number of lines of the vulnerable function. As shown in Fig.~\ref{fig:pacc_length}, PatchCon achieves the highest PACC in all bins. Both versions of the pairs with at most 20 lines contain at most 376 tokens and thus fit into a single 512-token chunk, so that MP does not affect their representations. On these pairs, PatchCon still achieves a PACC of 0.3636, compared with 0.3131 for DELTA and 0.2727 for SCL-CVD, indicating that its advantage stems from the patch-aware contrastive objective rather than from long-function coverage alone. As functions become longer, the PACC of the baselines decreases markedly: for functions with more than 80 lines, DELTA drops to 0.0901 and SCL-CVD to 0.0541, whereas PatchCon remains at 0.3063. Since baselines such as SCL-CVD and DELTA truncate each function to 400 and 512 tokens, respectively, vulnerability-relevant statements in long functions may be lost, whereas PatchCon covers up to 2,048 tokens through chunk-based encoding.

\begin{figure}[htbp]
\centering
\includegraphics[width=0.75\linewidth]{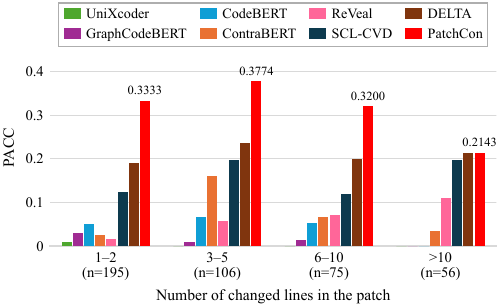}
\caption{PACC on the PairVul-Linux test set grouped by the number of changed lines in the patch; $n$ denotes the number of test pairs in each group, and the values above the bars give the PACC of PatchCon. ReVeal is evaluated on the pairs for which graphs can be constructed.}
\label{fig:pacc_patch}
\end{figure}

Finally, we group the test pairs by patch size, measured as the number of changed lines in a line-level diff between the vulnerable and patched versions. As shown in Fig.~\ref{fig:pacc_patch}, 195 of the 432 test pairs (45.1\%) are fixed by changing only one or two lines, so that the two versions are almost identical. On these pairs, PatchCon achieves a PACC of 0.3333, compared with 0.1897 for DELTA and 0.1231 for SCL-CVD. PatchCon also achieves the highest PACC on patches with 3--5 and 6--10 changed lines (0.3774 and 0.3200), whereas on patches with more than 10 changed lines, its PACC (0.2143) is on par with that of DELTA. This indicates that patch-aware hard negatives are most beneficial when the vulnerable and patched versions differ only slightly, which is the setting PatchCon is designed for.

\begin{answerbox}
\textbf{Answer to RQ4.} PatchCon improves patch-level discrimination across most CWE categories, although the magnitude of improvement varies by vulnerability type and is not explained by the number of training pairs. PatchCon also achieves the highest or tied-highest PACC across all function lengths and patch sizes, including short functions where MP has no effect, and its advantage is most pronounced on long functions and on small patches that change only a few lines.
\end{answerbox}

\subsection{Statistical Significance Analysis (RQ5)}
\label{subsec:statistical_reliability}

To assess the statistical significance of the observed improvements, we compute 95\% bootstrap confidence intervals for MCC and PACC. For MCC, bootstrap resampling is performed over function-level test samples, while for PACC, resampling is performed over patch pairs. For model-difference intervals, we perform paired bootstrap resampling on the same patch pairs to preserve sample-level correspondence between PatchCon and each baseline. We repeat the bootstrap procedure 1,000 times and report the 2.5th and 97.5th percentiles as the confidence interval. As shown in Table~\ref{tab:bootstrap_ci_compact}, PatchCon consistently achieves higher MCC and PACC confidence intervals than all baselines. More importantly, all $\Delta$PACC confidence intervals are strictly above zero, and so are the $\Delta$MCC confidence intervals for all baselines except DELTA, whose $\Delta$MCC interval ([-0.0075, +0.1256]) includes zero. This indicates that the improvements of PatchCon in vulnerable--patched discrimination are statistically significant under bootstrap resampling of the test set, whereas its MCC improvement over DELTA is not. The advantage is particularly clear on PACC, indicating that the proposed strategy substantially improves the model's ability to distinguish vulnerable functions from their patched counterparts.

\begin{table}[htbp]
\centering
\caption{Bootstrap 95\% confidence intervals (CIs) on key metrics. $\Delta$ denotes PatchCon minus the corresponding baseline.}
\label{tab:bootstrap_ci_compact}
\begin{adjustbox}{max width=\textwidth}
\begin{tabular}{lcccc}
\toprule
\textbf{Approach} & \textbf{MCC 95\% CI} & \textbf{$\Delta$MCC 95\% CI} & \textbf{PACC 95\% CI} & \textbf{$\Delta$PACC 95\% CI} \\
\midrule
UniXcoder     & [0.0000, 0.0779]  & [+0.0915, +0.2388] & [0.0000, 0.0116] & [+0.2778, +0.3681] \\
GraphCodeBERT & [-0.0212, 0.1075] & [+0.0850, +0.2333] & [0.0069, 0.0324] & [+0.2638, +0.3542] \\
CodeBERT      & [-0.0020, 0.1194] & [+0.0776, +0.2135] & [0.0278, 0.0718] & [+0.2315, +0.3241] \\
ContraBERT    & [-0.0294, 0.1019] & [+0.1025, +0.2391] & [0.0440, 0.0880] & [+0.2106, +0.3079] \\
ReVeal$^{\mathrm{a}}$ & [-0.0643, 0.0739] & [+0.1316, +0.2742] & [0.0288, 0.0697] & [+0.2307, +0.3341] \\
SCL-CVD       & [0.0395, 0.1721]  & [+0.0325, +0.1643] & [0.1157, 0.1829] & [+0.1273, +0.2292] \\
DELTA         & [0.0811, 0.2166]  & [-0.0075, +0.1256] & [0.1666, 0.2431] & [+0.0694, +0.1714] \\
\midrule
\textbf{PatchCon} & \textbf{[0.1456, 0.2710]} & -- & \textbf{[0.2824, 0.3704]} & -- \\
\bottomrule
\end{tabular}
\end{adjustbox}
\tabnotes{$^{\mathrm{a}}$Computed on the 416 test pairs available for ReVeal.}
\end{table}

\begin{answerbox}
\textbf{Answer to RQ5.} The strictly positive paired bootstrap confidence intervals for $\Delta$PACC indicate that PatchCon's improvements in vulnerable--patched discrimination over all baselines are statistically significant under bootstrap resampling of the test set; for MCC, the improvement is statistically significant over all baselines except DELTA.
\end{answerbox}

\subsection{Generalization to Future Vulnerabilities (RQ6)}
\label{subsec:ood}

To answer RQ6, we apply the models trained on PairVul-Linux in RQ1 to VentiVul without retraining, fine-tuning, or threshold adjustment. Since all CVEs in VentiVul were published more than a year after the most recent CVE in PairVul-Linux and none of them appears in PairVul-Linux, this evaluation is free from the CVE overlap discussed in Section~\ref{subsec:cve_overlap} and reflects how the models perform on vulnerabilities that were unknown at training time. Since RQ6 concerns vulnerable--patched discrimination, we evaluate the models on the 111 vulnerable--patched pairs with PACC, MCG, and the failure modes defined in Section~\ref{subsec:metrics}.

\begin{table}[htbp]
\centering
\caption{Pairwise results on VentiVul (111 vulnerable--patched pairs from 96 CVEs published in May 2025). P-V, P-B, and P-R are defined as in Table~\ref{tab:diagnostics}.}
\label{tab:ood_paired}
\begin{adjustbox}{max width=\textwidth}
\begin{tabular}{lccccc}
\toprule
\textbf{Approach} & \textbf{PACC} & \textbf{MCG} & \textbf{P-V} & \textbf{P-B} & \textbf{P-R} \\
\midrule
UniXcoder     & 0.0000 & 0.0034 & 1.0000 & 0.0000 & 0.0000 \\
GraphCodeBERT & 0.0000 & 0.0051 & 1.0000 & 0.0000 & 0.0000 \\
CodeBERT      & 0.0450 & 0.0068 & 0.8559 & 0.0811 & 0.0180 \\
ContraBERT    & 0.0270 & 0.0012 & 0.7838 & 0.1441 & 0.0450 \\
ReVeal$^{\mathrm{a}}$ & 0.0459 & -0.0001 & 0.7064 & 0.2018 & 0.0459 \\
SCL-CVD       & 0.1351 & 0.0278 & 0.4324 & 0.4054 & 0.0270 \\
DELTA         & 0.1261 & \textbf{0.0324} & 0.4865 & 0.3694 & 0.0180 \\
\midrule
\textbf{PatchCon} & \textbf{0.1622} & \textbf{0.0324} & 0.4324 & 0.2973 & 0.1081 \\
\bottomrule
\end{tabular}
\end{adjustbox}
\tabnotes{$^{\mathrm{a}}$Graph construction failed for 2 of the 111 pairs, so ReVeal is evaluated on the remaining 109 pairs.}
\end{table}

Table~\ref{tab:ood_paired} reports the results on VentiVul. Under this temporal distribution shift, all models perform worse than on the PairVul-Linux test set. The pretrained encoders predict almost all functions as vulnerable, so their PACC remains close to zero and most of their pairs fall into P-V. Among all models, PatchCon achieves the highest PACC (0.1622) and the highest MCG (0.0324, on par with DELTA), followed by SCL-CVD (0.1351) and DELTA (0.1261); the same ordering holds on all 116 pairs before removing the five pairs described in Section~\ref{subsec:datasets} (0.1552, 0.1293, and 0.1207). These results are consistent with those on unseen CVEs in Section~\ref{subsec:cve_overlap}, where PatchCon also achieves the highest PACC, and suggest that the vulnerable--patched discrimination learned by PatchCon transfers to newly disclosed vulnerabilities better than that of the compared models. Since VentiVul contains only 111 pairs, the differences among PatchCon, SCL-CVD, and DELTA should be interpreted with caution.

\begin{answerbox}
\textbf{Answer to RQ6.} On newly disclosed Linux kernel vulnerabilities, all models perform worse than on PairVul-Linux, while PatchCon still achieves the best vulnerable--patched discrimination among the evaluated models in terms of PACC and MCG.
\end{answerbox}

\section{Discussion}
\label{sec:discussion}

\subsection{Robustness to CVE Overlap}
\label{subsec:cve_overlap}
Following the original PairVul benchmark, our split is performed at the pair level, so different function pairs of the same CVE may appear in both the training and test sets. To examine whether the advantage of PatchCon depends on such overlap, we divide the test pairs into pairs whose CVE appears in the training set (\textit{Seen}, 271 pairs) and pairs whose CVE appears in neither the training nor the validation set (\textit{Unseen}, 154 pairs); the remaining seven test pairs share their CVE only with the validation set. Table~\ref{tab:cve_overlap} reports PACC and MCG on both subsets.

\begin{table}[htbp]
\centering
\caption{PACC and MCG on all test pairs, on test pairs whose CVE appears in the training set (Seen, 271 pairs), and on test pairs whose CVE appears in neither the training nor the validation set (Unseen, 154 pairs).}
\label{tab:cve_overlap}
\begin{adjustbox}{max width=\textwidth}
\begin{tabular}{lcccccc}
\toprule
\multirow{2}{*}{\textbf{Approach}} & \multicolumn{3}{c}{\textbf{PACC}} & \multicolumn{3}{c}{\textbf{MCG}} \\
\cmidrule(lr){2-4} \cmidrule(lr){5-7}
 & \textbf{All} & \textbf{Seen} & \textbf{Unseen} & \textbf{All} & \textbf{Seen} & \textbf{Unseen} \\
\midrule
UniXcoder          & 0.0046 & 0.0074 & 0.0000 & 0.0024 & 0.0025 & 0.0020 \\
GraphCodeBERT      & 0.0185 & 0.0295 & 0.0000 & 0.0076 & 0.0117 & 0.0015 \\
CodeBERT           & 0.0486 & 0.0554 & 0.0325 & 0.0035 & 0.0036 & 0.0033 \\
ContraBERT         & 0.0671 & 0.0812 & 0.0455 & 0.0026 & 0.0027 & 0.0025 \\
ReVeal$^{\mathrm{a}}$ & 0.0481 & 0.0467 & 0.0395 & 0.0000 & 0.0000 & 0.0001 \\
SCL-CVD            & 0.1505 & 0.1624 & 0.1169 & 0.0271 & 0.0325 & 0.0173 \\
DELTA              & 0.2060 & 0.2325 & 0.1558 & 0.0855 & 0.1104 & 0.0383 \\
\midrule
\textbf{PatchCon}  & \textbf{0.3264} & \textbf{0.3727} & \textbf{0.2403} & \textbf{0.1838} & \textbf{0.2442} & \textbf{0.0828} \\
\bottomrule
\end{tabular}
\end{adjustbox}
\tabnotes{$^{\mathrm{a}}$Evaluated on the test pairs available for ReVeal (257 seen and 152 unseen pairs).}
\end{table}

PatchCon achieves the highest PACC and MCG on both subsets. On unseen CVEs, its PACC (0.2403) is 54\% and 106\% higher than that of DELTA (0.1558) and SCL-CVD (0.1169), respectively, and its MCG (0.0828) is more than twice that of DELTA (0.0383). All models obtain lower scores on unseen CVEs, for which no pair of the same CVE is available during training, but the ranking of the three strongest models remains unchanged. The components analyzed in RQ2 also remain effective on unseen CVEs: removing patch pairs reduces PACC from 0.2403 to 0.0844, removing MP reduces it to 0.1234, and patch-aware weighting improves it over the unweighted SupCon variant from 0.1818 to 0.2403. These results indicate that the advantage of PatchCon in vulnerable--patched discrimination persists without CVE overlap, consistent with its results on newly disclosed vulnerabilities in RQ6. Since pair-level splits are also used in the original PairVul benchmark~\citep{du26vulrag}, we recommend that future evaluations on paired vulnerability benchmarks additionally report results on CVE-disjoint test pairs.

\subsection{Case Study}

\begin{figure}[htbp]
\centering
\includegraphics[width=1\linewidth]{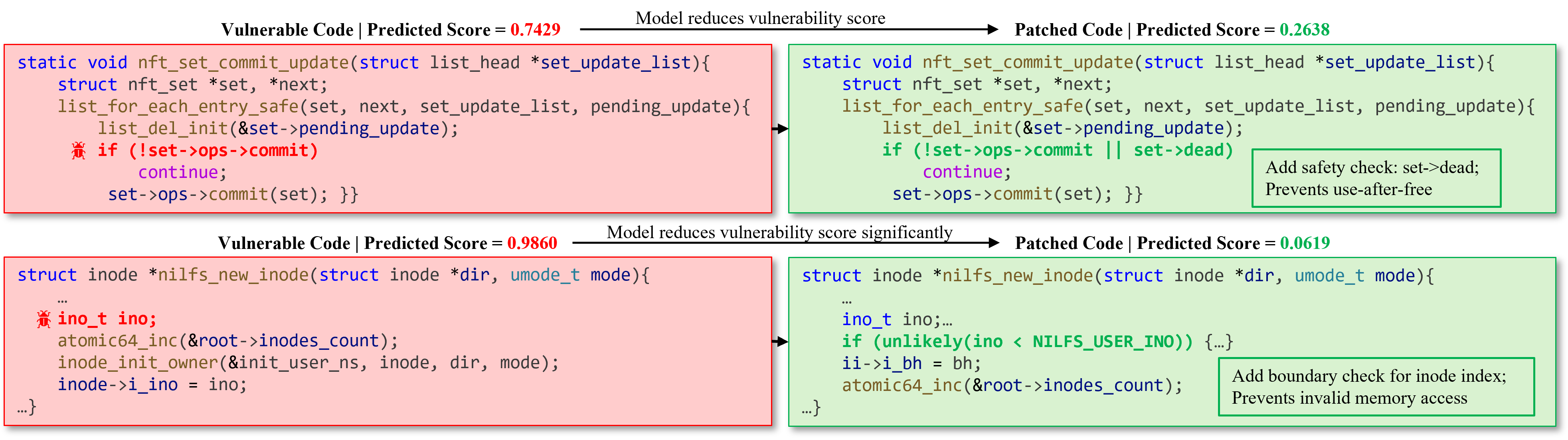}
\caption{Case studies of PatchCon. Case 1 (up): CVE-2024-0193 (CWE-416), Case 2 (down): CVE-2022-3649 (CWE-119).}
\label{fig:case_study}
\end{figure}

We further analyze three real-world vulnerability cases from the PairVul-Linux test set, including two correctly distinguished pairs and one failure case, to examine which security-relevant patch differences PatchCon captures and where it fails. None of the three CVEs appears in the training or validation data.
For each case, we compare a vulnerable function and its patched version, together with their predicted vulnerability scores, providing intuitive evidence of whether PatchCon captures subtle but security-critical differences introduced by patches. Fig.~\ref{fig:case_study} shows the two correctly distinguished examples from CWE-416 and CWE-119. In both cases, PatchCon assigns substantially higher vulnerability scores to the vulnerable functions than to their patched counterparts. For case 1, the vulnerable function receives a score of 0.7429, while the patched version is assigned a lower score of 0.2638. The key difference lies in the additional condition ``\texttt{|| set->dead}'' introduced in the patched version, which prevents operations on invalid or freed objects; PatchCon successfully captures this semantic change and reduces the vulnerability confidence accordingly.

A similar pattern is observed in case 2, where the vulnerable function receives a very high score of 0.9860, while the patched version is assigned a much lower score of 0.0619. The patch introduces additional boundary checks for inode allocation, ensuring that invalid inode indices are filtered out before further operations. Although the two functions are highly similar at the syntactic level, PatchCon distinguishes them after the safety checks are introduced, which is consistent with sensitivity to fine-grained vulnerability-fixing semantics. These examples suggest that, in such cases, the model responds to patch-level differences related to vulnerability causes and fixes rather than only to global code patterns. The consistent separation between vulnerable and patched functions in terms of prediction scores also aligns with the improvements observed in PACC and MCG, providing qualitative support for the quantitative results.

However, PatchCon does not always capture such checks. In another case (CVE-2022-3113, CWE-476), the patch adds a missing NULL check after a memory allocation in \texttt{mtk\_\allowbreak vcodec\_\allowbreak fw\_\allowbreak vpu\_\allowbreak init()}, but PatchCon assigns almost identical scores to both versions (0.9981 and 0.9988). Since allocation-and-dereference patterns are common in kernel code regardless of whether a guard is present, such a small defensive check may not change the function representation sufficiently, which suggests a direction for future improvement.

Overall, these case studies suggest that PatchCon assigns clearly lower scores to patched functions when security-relevant checks are introduced, and the resulting score differences may assist security developers in reviewing whether a patch reduces the predicted vulnerability risk.

\subsection{Limitations}
Although PatchCon improves both standard vulnerability detection and patch-pair discrimination metrics, several limitations remain. PatchCon relies on the availability of vulnerable--patched function pairs, which may be noisy or incomplete in projects beyond the Linux kernel, and its hyperparameters are tuned on the Linux kernel, so the optimal weighting may differ across projects. PatchCon targets function-level detection and does not localize vulnerable statements; moreover, since our evaluation focuses on vulnerable functions and their patched counterparts, its false-positive behavior on ordinary functions in file- or repository-level scanning remains to be evaluated. Per-CWE results are limited by the small number of test pairs per CWE type. Finally, our main experiments follow the pair-level split of PairVul; although the analyses on unseen CVEs and on VentiVul show that the advantage of PatchCon persists, function-level classification on unseen CVEs remains challenging for all models, and a CVE-grouped split would provide a stricter evaluation. Future work may explore richer patch representations, line-level explanations, hierarchical CWE-aware contrastive objectives, and stronger code encoders to improve robustness and interpretability.

\subsection{Threats to Validity}
Following \citet{wohlin2012experimentation}, we discuss threats to construct, internal, external, and conclusion validity, together with the strategies used to mitigate them.

\textbf{Construct validity.} Standard metrics may not reflect vulnerability-fixing semantics; thus, we include PACC and MCG, which directly measure vulnerable--patched discrimination, while acknowledging that they remain proxies.

\textbf{Internal validity.} Differences in model architectures and training setups may affect the comparison between PatchCon and the baselines. We reduce this threat by using consistent data splits and evaluation protocols and clearly distinguishing reproduced and non-reproduced results. In addition, the pair-level split allows pairs from the same CVE to appear in both the training and test sets, which may lead to optimistic performance estimates. To mitigate this threat, we additionally evaluate all models on test pairs whose CVE appears in neither the training nor the validation set (Section~\ref{subsec:cve_overlap}) and on newly disclosed vulnerabilities (RQ6).

\textbf{External validity.} Our findings may be affected by project-specific bias, as our experiments use Linux kernel vulnerabilities. The Linux kernel is a large and security-critical C codebase with a large number of disclosed vulnerabilities and fixes, but our findings may not fully generalize to other projects. In addition, our evaluation is constrained to C code; future work should extend evaluation to more languages, datasets, and industrial deployment scenarios.

\textbf{Conclusion validity.} Randomness in training may influence the results. The bootstrap confidence intervals and model-difference confidence intervals quantify the uncertainty due to test-set sampling, but each configuration is trained once with a fixed random seed, so the variance across training runs is not captured.

\section{Conclusion}
\label{sec:conclusion}
In this paper, we proposed PatchCon, a patch-aware supervised contrastive learning framework for function-level vulnerability detection. PatchCon explicitly leverages vulnerable--patched function pairs as hard negatives and combines chunk-based max pooling, classification learning, and weighted contrastive learning. Experiments on PairVul-Linux show that PatchCon improves both standard detection metrics and pairwise discrimination metrics, and its gains in pairwise accuracy over all evaluated baselines are statistically significant. Ablation studies, hyperparameter sensitivity analysis, backbone comparisons, fine-grained analyses by CWE type, function length, and patch size, and case studies further support the effectiveness of the proposed design. Among the reproduced models, PatchCon also achieves the highest pairwise accuracy on test pairs from CVEs unseen during training and on Linux kernel vulnerabilities disclosed after the training data. Our analysis further suggests that future evaluations on paired vulnerability benchmarks should also report results on CVE-disjoint test pairs.

\begin{declitem}{Data availability}The processed datasets and splits, the source code of PatchCon, the trained models, the per-sample predictions of PatchCon, and the scripts used for the analyses in this paper are available at \url{https://figshare.com/s/cb1d02d2cea2f6582744}.\end{declitem}
\begin{declitem}{Funding}Chaomeng Lu is supported by the CSC-IMEC-KU Leuven scholarship, jointly funded by the China Scholarship Council (CSC) and KU Leuven (CSC No. 202308130074).\end{declitem}
\begin{declitem}{Author contributions}Chaomeng Lu was responsible for conceptualization, methodology, software, data curation, experiments, analysis, visualization, and writing of the original draft. Pooya Rostami Mazrae contributed to the conceptualization of the study and reviewed and edited the manuscript. Bert Lagaisse supervised the work and reviewed and edited the manuscript. \end{declitem}

\section*{Declarations}
\begin{declitem}{Ethical Approval}Not applicable.\end{declitem}
\begin{declitem}{Informed Consent}Not applicable.\end{declitem}
\begin{declitem}{Conflict of interest}The authors declare that they have no conflict of interest.\end{declitem}
\begin{declitem}{Clinical Trial Number}Not applicable.\end{declitem}

\bibliographystyle{spbasic}
\bibliography{reference}

@INPROCEEDINGS{tan25similar,
  author={Tan, Zixuan and Zhou, Jiayuan and Hu, Xing and Pan, Shengyi and Liu, Kui and Xia, Xin},
  booktitle={2025 IEEE/ACM 47th International Conference on Software Engineering (ICSE)}, 
  title={Similar but Patched Code Considered Harmful: The Impact of Similar but Patched Code on Recurring Vulnerability Detection and How to Remove Them}, 
  year={2025},
  volume={},
  number={},
  pages={2445-2457},
  doi={10.1109/ICSE55347.2025.00110}}

@article{croft2022data,
  title={Data preparation for software vulnerability prediction: A systematic literature review},
  author={Croft, Roland and Xie, Yongzheng and Babar, Muhammad Ali},
  journal={IEEE Trans Softw Eng},
  volume={49},
  number={3},
  pages={1044--1063},
  year={2022},
  doi={10.1109/TSE.2022.3171202},
  publisher={IEEE}
}

@article{le2022survey,
  title={A survey on data-driven software vulnerability assessment and prioritization},
  author={Le, Triet H. M. and Chen, Huaming and Babar, M Ali},
  journal={ACM Comput Surv},
  volume={55},
  number={5},
  pages={1--39},
  year={2022},
  doi={10.1145/3529757},
  publisher={ACM New York, NY}
}

@INPROCEEDINGS{steenhoek23an,
  author={Steenhoek, Benjamin and Rahman, Md Mahbubur and Jiles, Richard and Le, Wei},
  booktitle={2023 IEEE/ACM 45th International Conference on Software Engineering (ICSE)}, 
  title={An Empirical Study of Deep Learning Models for Vulnerability Detection}, 
  year={2023},
  volume={},
  number={},
  pages={2237-2248},
  doi={10.1109/ICSE48619.2023.00188}}

@article{zhou25large,
author = {Zhou, Xin and Cao, Sicong and Sun, Xiaobing and Lo, David},
title = {Large Language Model for Vulnerability Detection and Repair: Literature Review and the Road Ahead},
year = {2025},
issue_date = {June 2025},
publisher = {Association for Computing Machinery},
address = {New York, NY, USA},
volume = {34},
number = {5},
issn = {1049-331X},
doi = {10.1145/3708522},
journal = {ACM Trans Softw Eng Methodol},
month = may,
articleno = {145},
numpages = {31}
}

@inproceedings{ladisa2023sok,
  title={{SoK}: Taxonomy of attacks on open-source software supply chains},
  author={Ladisa, Piergiorgio and Plate, Henrik and Martinez, Matias and Barais, Olivier},
  booktitle={2023 IEEE Symposium on Security and Privacy (SP)},
  pages={1509--1526},
  year={2023},
  doi={10.1109/SP46215.2023.10179304},
  organization={IEEE}
}

@inproceedings{ohm2020backstabber,
  title={Backstabber’s knife collection: A review of open source software supply chain attacks},
  author={Ohm, Marc and Plate, Henrik and Sykosch, Arnold and Meier, Michael},
  booktitle={International Conference on Detection of Intrusions and Malware, and Vulnerability Assessment},
  pages={23--43},
  year={2020},
  doi={10.1007/978-3-030-52683-2_2},
  organization={Springer}
}

@inproceedings {niklas24uncover,
author = {Niklas Risse and Marcel B{\"o}hme},
title = {Uncovering the Limits of Machine Learning for Automatic Vulnerability Detection},
booktitle = {33rd USENIX Security Symposium (USENIX Security 24)},
year = {2024},
isbn = {978-1-939133-44-1},
address = {Philadelphia, PA},
pages = {4247--4264},
publisher = {USENIX Association},
month = aug
}

@article{lu2025from,
  title={From Lab to Reality: A Practical Evaluation of Deep Learning Models and {LLMs} for Vulnerability Detection},
  author={Lu, Chaomeng and Lagaisse, Bert},
  journal={J Syst Softw},
  year={2026},
  note={(in press). Preprint available at \url{https://doi.org/10.48550/arXiv.2512.10485}}
}

@INPROCEEDINGS{croft23data,
  author={Croft, Roland and Babar, M. Ali and Kholoosi, M. Mehdi},
  booktitle={2023 IEEE/ACM 45th International Conference on Software Engineering (ICSE)}, 
  title={Data Quality for Software Vulnerability Datasets}, 
  year={2023},
  volume={},
  number={},
  pages={121-133},
  doi={10.1109/ICSE48619.2023.00022}}

@ARTICLE{chakraborty24revi,
  author={Chakraborty, Partha and Arumugam, Krishna Kanth and Alfadel, Mahmoud and Nagappan, Meiyappan and McIntosh, Shane},
  journal={IEEE Trans Softw Eng}, 
  title={Revisiting the Performance of Deep Learning-Based Vulnerability Detection on Realistic Datasets}, 
  year={2024},
  volume={50},
  number={8},
  pages={2163-2177},
  doi={10.1109/TSE.2024.3423712}}

@ARTICLE{wen24livable,
  author={Wen, Xin-Cheng and Gao, Cuiyun and Luo, Feng and Wang, Haoyu and Li, Ge and Liao, Qing},
  journal={IEEE Trans Softw Eng}, 
  title={{LIVABLE}: Exploring Long-Tailed Classification of Software Vulnerability Types}, 
  year={2024},
  volume={50},
  number={6},
  pages={1325-1339},
  doi={10.1109/TSE.2024.3382361}}

@article{shiri24a,
author = {Shiri Harzevili, Nima and Boaye Belle, Alvine and Wang, Junjie and Wang, Song and Jiang, Zhen Ming and Nagappan, Nachiappan},
title = {A Systematic Literature Review on Automated Software Vulnerability Detection Using Machine Learning},
year = {2024},
issue_date = {March 2025},
publisher = {Association for Computing Machinery},
address = {New York, NY, USA},
volume = {57},
number = {3},
issn = {0360-0300},
doi = {10.1145/3699711},
journal = {ACM Comput Surv},
month = nov,
articleno = {55},
numpages = {36}
}

@INPROCEEDINGS{yamaguchi14modeling,
  author={Yamaguchi, Fabian and Golde, Nico and Arp, Daniel and Rieck, Konrad},
  booktitle={2014 IEEE Symposium on Security and Privacy}, 
  title={Modeling and Discovering Vulnerabilities with Code Property Graphs}, 
  year={2014},
  volume={},
  number={},
  pages={590-604},
  doi={10.1109/SP.2014.44}}

@article{ghaffarian18software,
author = {Ghaffarian, Seyed Mohammad and Shahriari, Hamid Reza},
title = {Software Vulnerability Analysis and Discovery Using Machine-Learning and Data-Mining Techniques: A Survey},
year = {2017},
issue_date = {July 2018},
publisher = {Association for Computing Machinery},
address = {New York, NY, USA},
volume = {50},
number = {4},
issn = {0360-0300},
doi = {10.1145/3092566},
journal = {ACM Comput Surv},
month = aug,
articleno = {56},
numpages = {36}
}

@article{sheng26large,
author = {Sheng, Ze and Chen, Zhicheng and Gu, Shuning and Huang, Heqing and Gu, Guofei and Huang, Jeff},
title = {{LLMs} in Software Security: A Survey of Vulnerability Detection Techniques and Insights},
year = {2025},
issue_date = {April 2026},
publisher = {Association for Computing Machinery},
address = {New York, NY, USA},
journal = {ACM Comput Surv},
volume = {58},
number = {5},
issn = {0360-0300},
doi = {10.1145/3769082},
month = nov,
articleno = {134},
numpages = {35}
}

@article{cao2025mcl,
  title={{MCL-VD}: Multi-modal contrastive learning with {LoRA}-enhanced {GraphCodeBERT} for effective vulnerability detection},
  author={Cao, Yi and Ju, Xiaolin and Chen, Xiang and Gong, Lina},
  journal={Autom Softw Eng},
  volume={32},
  number={2},
  pages={67},
  year={2025},
  doi={10.1007/s10515-025-00543-3},
  publisher={Springer}
}

@inproceedings{ji24applying,
    title = "Applying Contrastive Learning to Code Vulnerability Type Classification",
    author = "Ji, Chen  and
      Yang, Su  and
      Sun, Hongyu  and
      Zhang, Yuqing",
    editor = "Al-Onaizan, Yaser  and
      Bansal, Mohit  and
      Chen, Yun-Nung",
    booktitle = "Proceedings of the 2024 Conference on Empirical Methods in Natural Language Processing",
    month = nov,
    year = "2024",
    address = "Miami, Florida, USA",
    publisher = "Association for Computational Linguistics",
    doi = "10.18653/v1/2024.emnlp-main.666",
    pages = "11942--11952"
}

@inproceedings{chen24clear,
author = {Chen, Yizhou and Sun, Zeyu and Gong, Zhihao and Hao, Dan},
title = {Improving Smart Contract Security with Contrastive Learning-based Vulnerability Detection},
year = {2024},
isbn = {9798400702174},
publisher = {Association for Computing Machinery},
address = {New York, NY, USA},
doi = {10.1145/3597503.3639173},
booktitle = {Proceedings of the IEEE/ACM 46th International Conference on Software Engineering},
articleno = {156},
numpages = {11},
location = {Lisbon, Portugal},
series = {ICSE '24}
}

@article{xu23contrast,
author = {Xu, Lingling and Xie, Haoran and Li, Zongxi and Wang, Fu Lee and Wang, Weiming and Li, Qing},
title = {Contrastive Learning Models for Sentence Representations},
year = {2023},
issue_date = {August 2023},
publisher = {Association for Computing Machinery},
address = {New York, NY, USA},
volume = {14},
number = {4},
issn = {2157-6904},
doi = {10.1145/3593590},
journal = {ACM Trans Intell Syst Technol},
month = jun,
articleno = {67},
numpages = {34}
}

@inproceedings{chen22path,
author = {Cheng, Xiao and Zhang, Guanqin and Wang, Haoyu and Sui, Yulei},
title = {Path-sensitive code embedding via contrastive learning for software vulnerability detection},
year = {2022},
isbn = {9781450393799},
publisher = {Association for Computing Machinery},
address = {New York, NY, USA},
doi = {10.1145/3533767.3534371},
booktitle = {Proceedings of the 31st ACM SIGSOFT International Symposium on Software Testing and Analysis},
pages = {519--531},
numpages = {13},
location = {Virtual, South Korea},
series = {ISSTA 2022}
}

@article{sun21vdsimilar,
title = {{VDSimilar}: Vulnerability detection based on code similarity of vulnerabilities and patches},
journal = {Comput Secur},
volume = {110},
pages = {102417},
year = {2021},
issn = {0167-4048},
doi = {10.1016/j.cose.2021.102417},
author = {Hao Sun and Lei Cui and Lun Li and Zhenquan Ding and Zhiyu Hao and Jiancong Cui and Peng Liu}
}

@INPROCEEDINGS{shi23cocosoda,
  author={Shi, Ensheng and Wang, Yanlin and Gu, Wenchao and Du, Lun and Zhang, Hongyu and Han, Shi and Zhang, Dongmei and Sun, Hongbin},
  booktitle={2023 IEEE/ACM 45th International Conference on Software Engineering (ICSE)}, 
  title={{CoCoSoDa}: Effective Contrastive Learning for Code Search}, 
  year={2023},
  volume={},
  number={},
  pages={2198-2210},
  doi={10.1109/ICSE48619.2023.00185}}

@article{tian26when,
title = {When fixes teach: Repair-aware contrastive learning for optimization-resilient binary vulnerability detection},
journal = {J Syst Archit},
volume = {173},
pages = {103722},
year = {2026},
issn = {1383-7621},
doi = {10.1016/j.sysarc.2026.103722},
author = {Zhenzhou Tian and Jiale Zhao and Ming Fan and Jiaze Sun and Yanping Chen and Lingwei Chen}
}

@ARTICLE{chakraborty22deep,
  author={Chakraborty, Saikat and Krishna, Rahul and Ding, Yangruibo and Ray, Baishakhi},
  journal={IEEE Trans Softw Eng}, 
  title={Deep Learning Based Vulnerability Detection: Are We There Yet?}, 
  year={2022},
  volume={48},
  number={9},
  pages={3280-3296},
  doi={10.1109/TSE.2021.3087402}}

@article{wang24sclcvd,
title = {{SCL-CVD}: Supervised contrastive learning for code vulnerability detection via {GraphCodeBERT}},
journal = {Comput Secur},
volume = {145},
pages = {103994},
year = {2024},
issn = {0167-4048},
doi = {10.1016/j.cose.2024.103994},
author = {Rongcun Wang and Senlei Xu and Yuan Tian and Xingyu Ji and Xiaobing Sun and Shujuang Jiang}
}

@inproceedings{guo2021graphcb,
title={{GraphCodeBERT}: Pre-training Code Representations with Data Flow},
author={Daya Guo and Shuo Ren and Shuai Lu and Zhangyin Feng and Duyu Tang and Shujie Liu and Long Zhou and Nan Duan and Alexey Svyatkovskiy and Shengyu Fu and Michele Tufano and Shao Kun Deng and Colin Clement and Dawn Drain and Neel Sundaresan and Jian Yin and Daxin Jiang and Ming Zhou},
booktitle={International Conference on Learning Representations},
year={2021}
}

@InProceedings{chen20apmlr,
  title = 	 {A Simple Framework for Contrastive Learning of Visual Representations},
  author =       {Chen, Ting and Kornblith, Simon and Norouzi, Mohammad and Hinton, Geoffrey},
  booktitle = 	 {Proceedings of the 37th International Conference on Machine Learning},
  pages = 	 {1597--1607},
  year = 	 {2020},
  editor = 	 {Daum{\'e}, III, Hal and Singh, Aarti},
  volume = 	 {119},
  series = 	 {Proceedings of Machine Learning Research},
  month = 	 {13--18 Jul},
  publisher =    {PMLR}
}

@InProceedings{He2020CVPR,
author = {He, Kaiming and Fan, Haoqi and Wu, Yuxin and Xie, Saining and Girshick, Ross},
title = {Momentum Contrast for Unsupervised Visual Representation Learning},
booktitle = {Proceedings of the IEEE/CVF Conference on Computer Vision and Pattern Recognition (CVPR)},
month = {June},
year = {2020},
pages = {9726--9735},
doi = {10.1109/CVPR42600.2020.00975}
}

@INPROCEEDINGS{contrabert23,
  author={Liu, Shangqing and Wu, Bozhi and Xie, Xiaofei and Meng, Guozhu and Liu, Yang},
  booktitle={2023 IEEE/ACM 45th International Conference on Software Engineering (ICSE)}, 
  title={{ContraBERT}: Enhancing Code Pre-trained Models via Contrastive Learning}, 
  year={2023},
  volume={},
  number={},
  pages={2476-2487},
  doi={10.1109/ICSE48619.2023.00207}}

@inproceedings{codebert20,
  author    = {Zhangyin Feng and Daya Guo and Duyu Tang and Nan Duan and Xiaocheng Feng and Ming Gong and Linjun Shou and Bing Qin and Ting Liu and Daxin Jiang and Ming Zhou},
  title     = {{CodeBERT}: A Pre-Trained Model for Programming and Natural Languages},
  booktitle = {Findings of the Association for Computational Linguistics: EMNLP 2020},
  year      = {2020},
  pages     = {1536--1547},
  publisher = {Association for Computational Linguistics},
  doi       = {10.18653/v1/2020.findings-emnlp.139}
}

@inproceedings{unixcoder22,
  author    = {Daya Guo and Shuai Lu and Nan Duan and Yanlin Wang and Ming Zhou and Jian Yin},
  title     = {{UniXcoder}: Unified Cross-Modal Pre-training for Code Representation},
  booktitle = {Proceedings of the 60th Annual Meeting of the Association for Computational Linguistics},
  year      = {2022},
  publisher = {Association for Computational Linguistics},
  pages     = {7212--7225},
  doi       = {10.18653/v1/2022.acl-long.499}
}

@inproceedings{supcon20,
  author    = {Prannay Khosla and Piotr Teterwak and Chen Wang and Aaron Sarna and Yonglong Tian and Phillip Isola and Aaron Maschinot and Ce Liu and Dilip Krishnan},
  title     = {Supervised Contrastive Learning},
  booktitle = {Advances in Neural Information Processing Systems},
  volume    = {33},
  year      = {2020}
}

@INPROCEEDINGS {primevul25,
author = { Ding, Yangruibo and Fu, Yanjun and Ibrahim, Omniyyah and Sitawarin, Chawin and Chen, Xinyun and Alomair, Basel and Wagner, David and Ray, Baishakhi and Chen, Yizheng },
booktitle = { 2025 IEEE/ACM 47th International Conference on Software Engineering (ICSE) },
title = {Vulnerability Detection with Code Language Models: How Far Are We?},
year = {2025},
volume = {},
ISSN = {},
pages = {1729-1741},
doi = {10.1109/ICSE55347.2025.00038},
publisher = {IEEE Computer Society},
address = {Los Alamitos, CA, USA},
month =May
}

@article{du26vulrag,
author = {Du, Xueying and Zheng, Geng and Wang, Kaixin and Zou, Yi and Wang, Yujia and Deng, Wentai and Feng, Jiayi and Liu, Mingwei and Chen, Bihuan and Peng, Xin and Ma, Tao and Lou, Yiling},
title = {{Vul-RAG}: Enhancing {LLM}-based Vulnerability Detection via Knowledge-level {RAG}},
year = {2026},
issue_date = {October 2026},
publisher = {Association for Computing Machinery},
address = {New York, NY, USA},
volume = {35},
number = {10},
issn = {1049-331X},
doi = {10.1145/3797277},
journal = {ACM Trans Softw Eng Methodol},
month = sep,
articleno = {312},
numpages = {25}
}

@inproceedings{devign19,
  author    = {Yaqin Zhou and Shangqing Liu and Jingkai Siow and Xiaoning Du and Yang Liu},
  title     = {{Devign}: Effective Vulnerability Identification by Learning Comprehensive Program Semantics via Graph Neural Networks},
  booktitle = {Advances in Neural Information Processing Systems},
  volume    = {32},
  year      = {2019}
}

@inproceedings{vuldeepecker18,
  author    = {Zhen Li and Deqing Zou and Shouhuai Xu and Xinyu Ou and Hai Jin and Sujuan Wang and Zhijun Deng and Yuyi Zhong},
  title     = {{VulDeePecker}: A Deep Learning-Based System for Vulnerability Detection},
  booktitle = {Proceedings of the Network and Distributed System Security Symposium},
  series    = {NDSS},
  year      = {2018},
  doi       = {10.14722/ndss.2018.23158}
}

@article{sysevr22,
  author  = {Zhen Li and Deqing Zou and Shouhuai Xu and Hai Jin and Yawei Zhu and Zhaoxuan Chen},
  title   = {{SySeVR}: A Framework for Using Deep Learning to Detect Software Vulnerabilities},
  journal = {IEEE Trans Dependable Secure Comput},
  volume  = {19},
  number  = {4},
  pages   = {2244--2258},
  year    = {2022},
  doi     = {10.1109/TDSC.2021.3051525}
}

@inproceedings{chen23diverse,
author = {Chen, Yizheng and Ding, Zhoujie and Alowain, Lamya and Chen, Xinyun and Wagner, David},
title = {{DiverseVul}: A New Vulnerable Source Code Dataset for Deep Learning Based Vulnerability Detection},
year = {2023},
isbn = {9798400707650},
publisher = {Association for Computing Machinery},
address = {New York, NY, USA},
doi = {10.1145/3607199.3607242},
booktitle = {Proceedings of the 26th International Symposium on Research in Attacks, Intrusions and Defenses},
pages = {654--668},
numpages = {15},
location = {Hong Kong, China},
series = {RAID '23}
}

@book{wohlin2012experimentation,
  title     = {Experimentation in software engineering},
  author    = {Wohlin, Claes and Runeson, Per and H{\"o}st, Martin and Ohlsson, Magnus C. and Regnell, Bj{\"o}rn and Wessl{\'e}n, Anders},
  publisher = {Springer},
  address   = {Berlin, Heidelberg},
  year      = {2012},
  doi       = {10.1007/978-3-642-29044-2}
}

\end{document}